\documentclass[usenatbib,usegraphicx]{mn2e}
\usepackage{amsmath}

\usepackage[
pdfauthor={Laszlo Szucs},
pdftitle={The 12CO/13CO ratio in MCs},
pdfstartview=FitH,
pdfkeywords={astrochemistry} {hydrodynamics} {radiative transfer} {ISM: abundances},
linkcolor=blue,
anchorcolor=blue,
citecolor=blue,
filecolor=blue,
menucolor=blue,
urlcolor=blue,
unicode=true,
colorlinks=true]{hyperref}

\newcommand{\corr}[1]{{\textcolor{black}{#1}}}

\title[The $^{12}\textrm{CO}/^{13}\textrm{CO}$ ratio in MCs]{The $^{12}\textrm{CO}/^{13}\textrm{CO}$ ratio in turbulent molecular clouds}
\author[Sz\H{u}cs et al.]{L\'{a}szl\'{o} Sz\H{u}cs\thanks{Member of IMPRS for Astronomy \& Cosmic Physics at the
University of Heidelberg.}\thanks{szucsl@uni-heidelberg.de}, 
       Simon C.~O. Glover \& Ralf S. Klessen 
       \\ Universit\"{a}t Heidelberg, Zentrum f\"{u}r Astronomie, 
       Institut f\"{u}r Theoretische Astrophysik \\ Albert-Ueberle-Str. 2, D-69120 
       Heidelberg, Germany}

\newenvironment{itemize*}
  {\begin{itemize}
    \setlength{\itemsep}{0pt}
    \setlength{\parskip}{0pt}}
  {\end{itemize}}

\begin{document}

\maketitle

\begin{abstract}

\corr{
The $^{13}\textrm{CO}$ molecule is often used as a column density tracer in regions where the $^{12}\textrm{CO}$ emission saturates. The $^{13}\textrm{CO}$ column density is then related to that of $^{12}\textrm{CO}$ by a uniform isotopic ratio. A similar approximation is frequently used when deriving  $^{13}\textrm{CO}$ emission maps from numerical simulations of molecular clouds. 
To test this assumption we calculate the $^{12}\textrm{CO}/^{13}\textrm{CO}$ ratio self-consistently, taking the isotope selective photodissociation and the chemical fractionation of CO into account. We model the coupled chemical, thermal and dynamical evolution and the emergent $^{13}\textrm{CO}$ emission of isolated, starless molecular clouds in various environments. 
Selective photodissociation has a minimal effect on the ratio, while the chemical fractionation causes a factor of 2-3 decrease at intermediate cloud depths. The variation correlates with both the $^{12}\textrm{CO}$ and the $^{13}\textrm{CO}$ column densities. Neglecting the depth dependence results in $\leq$60 per cent error in $^{12}\textrm{CO}$ column densities derived from $^{13}\textrm{CO}$. The same assumption causes $\leq$50 per cent disparity in the $^{13}\textrm{CO}$ emission derived from simulated clouds. We show that the discrepancies can be corrected by a fitting formula. The formula is consistent with millimetre-wavelength isotopic ratio measurements of dense molecular clouds, but underestimates the ratios from the ultraviolet absorption of diffuse regions.
}

\end{abstract}


\section{Introduction}
The carbon monoxide molecule (CO) and its isotopes are the most widely used gas-phase tracers of total column density in the interstellar medium (ISM). In contrast to the hydrogen molecule, CO is asymmetric and hence has a permanent dipole moment. Its dipole transitions between rotational levels can be excited at temperatures (few$\times10$ K) and densities ($\approx$300 cm$^{-3}$) typical of giant molecular clouds (GMCs). The emission from the lowest transitions is relatively easily detectable at millimetre wavelengths. Due to its high fractional abundance ($\chi_{\rmn{CO}}\approx10^{-4}$, in equilibrium at high density) the emission of the most abundant CO isotope ($^{12}\textrm{CO}$) is usually optically thick, therefore only lower limits of the total column density could be derived. To achieve better total column density estimates, less abundant CO isotopes are used (usually $^{13}\textrm{CO}$ and $\textrm{C}^{18}\textrm{O}$). The simplest and most often used method is the following \citep[e.g.][]{Pineda2008,Wilson2009,Pineda2010}: The $^{12}\textrm{CO}$ emission is assumed to be fully optically thick and in local thermodynamic equilibrium (LTE), allowing the excitation temperature of $^{12}\textrm{CO}$ to be calculated. If the excitation temperature is the same for $^{13}\textrm{CO}$ and if $^{13}\textrm{CO}$ is optically thin, a simple relation between the integrated intensity along the line of sight, $W(^{13}$CO$)$, and the column density of $^{13}\textrm{CO}$, $N(^{13}\textrm{CO})$ can be derived \citep[e.g. see equation~9 in][]{Pineda2008}. The $^{13}\textrm{CO}$ column densities are then usually converted into $^{12}\textrm{CO}$ column densities using a uniform $^{12}\textrm{CO}/^{13}\textrm{CO}$ isotope ratio. Finally, the $^{12}\textrm{CO}$ column density is transformed to total column density assuming a given conversion factor between CO and H$_2$. The high uncertainties and environmental dependence of this final step have been extensively studied in the literature \citep[][and references within]{Shetty2011a,Shetty2011b,GloverMacLow2011,Feldmann2012}.

However, the $^{12}\textrm{CO}/^{13}\textrm{CO}$ ratio may vary considerably. Typically it is chosen to be equal to the measured $^{12}$C/$^{13}$C ratio \citep{Pineda2010}. The carbon isotope ratio, however, shows large regional variations. Based on observations of millimetre-wavelength emission of CO isotopes, \citet{LangerPenzias1990} found a systematic gradient with galactocentric distance in the carbon isotope ratio, ranging from 24 in the Galactic Center to about 70 at ${\sim}$12~kpc. They found the average ratio to be 57 at the solar galactocentric distance, which shows $^{13}$C enhancement compared to the value of 89 measured in the Solar System \citep{Geiss1988}. Observations of CO absorption in ultraviolet electronic and near-infrared vibrational transitions \citep[e.g.][]{Scoville1983,MitchellMaillard1993,Goto2003,Sonnentrucker2007,Sheffer2007} find up to a factor of 3 higher ratios in the solar neighbourhood. Nevertheless, the most frequently adopted ratios are between 57 \citep{LangerPenzias1990} and 69 \citep{Wilson1999}, the measured average values for \corr{the interstellar medium within a few kpc of the Sun}.

The gradient with galactocentric distance and the \corr{$^{13}$C enhancement in the solar neighbourhood compared to the Solar System value could} be interpreted in the framework of the carbon isotopic 
nucleosynthesis. $^{12}$C is the primary product of the triple-alpha process during the post Red 
Giant Branch (RGB) evolution of massive stars. The rarer $^{13}$C is produced from $^{12}$C as a 
secondary product in the CNO cycle during the RGB phase of low and intermediate mass stars. Due 
to the longer lifetime of low and intermediate mass stars -- which are the main contributors of 
$^{13}$C enrichment -- the $^{12}$C/$^{13}$C ratio is expected to decrease with time and to depend on the star formation history \citep{Audouze1975}.

\begin{figure}
\includegraphics[scale=0.42]{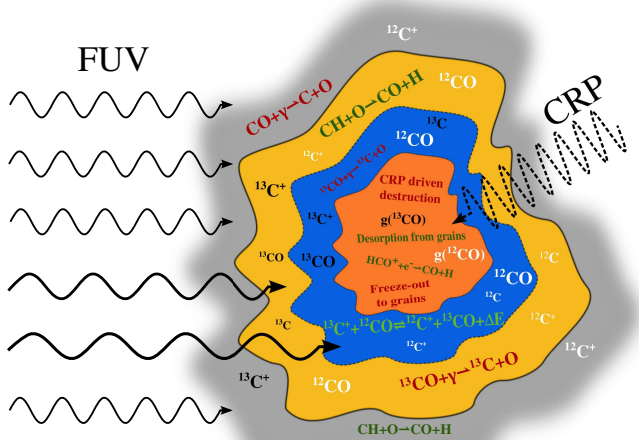} 
\caption{Qualitative picture of carbon isotopic chemistry in molecular clouds. The outer diffuse layer (gray) is followed by the translucent (yellow and blue) and the inner, dense regions (orange). The font size of the chemical species relates to the abundance of the species in the region. The font size of chemical reactions indicates the importance of the reaction in the region, the green and red colors represent reactions producing and destroying CO, respectively. When a particular isotopic species is not indicated, then both species are affected. Based on \citet{DishoeckBlack1988}.}
\label{fig:cochem}
\end{figure}

More important for our problem, however, is the fact that the CO isotope ratio could vary by a factor of a few even if the \corr{elemental} $^{12}$C/$^{13}$C ratio is constant in a region under investigation, due to isotope-selective chemical processes. For example, \citet{DishoeckBlack1988} describe qualitatively the CO isotopic chemistry in molecular clouds as follows (see Fig.~\ref{fig:cochem}). The preferred pathways of CO production are the ion-neutral reaction of C$^{+}$ and OH producing HCO$^{+}$, which dissociatively recombines to CO and H, and the neutral-neutral reaction of CH or CH$_{2}$ with an oxygen atom. These reactions are not isotope-selective and work (with varying efficiency) in every region of the molecular cloud. In the diffuse regions (gray, $A_{\rmn{V}} < 0.5\,\textrm{mag}$\footnote{The exact visual extinction values depend on the strength of the incident radiation field and the density distribution.}) the photodissociation of CO by interstellar far-ultraviolet (FUV) photons dominates over the production reactions and most of the carbon is in ionized form. 
\corr{In translucent regions (yellow, $1\,\textrm{mag} < A_{\rmn{V}} < 2\,\textrm{mag}$) the CO production rates start to compete with the photodissociation. The $^{12}\textrm{CO}$ column density becomes high enough to effectively self-shield itself from the incident interstellar FUV photons. However, $^{13}\textrm{CO}$, due to its slightly shifted absorption lines at UV wavelengths and the lower abundance, is less effectively self-shielded. This difference in self-shielding, in principle, leads to isotope-selective photodissociation. In practice, the selective photodissociation is expected to be dominant only in low density ($n < 10^{2} \textrm{cm}^{-3}$) domains or in dense regions with very strong radiation fields \citep[see][]{Roellig2013}.}
Further in (blue, $2\,\textrm{mag} < A_{\rmn{V}} < 5\,\textrm{mag}$), due to the shielding by dust absorption and the increasing  $^{13}\textrm{CO}$ column density, both isotopic species are effectively protected from FUV photons. In this region ionized carbon is still abundant and the 
\begin{equation}
{^{13}\rmn{C}^{+}} + {^{12}\rmn{CO}} \rightleftharpoons {^{12}\rmn{C}^{+}} + {^{13}\rmn{CO}} + \Delta\rmn{E} 
\label{eq:fracreact}
\end{equation}
fractionation reaction \citep{Watson1976} becomes important. At temperatures typical to the corresponding cloud depths, the exothermic reaction (to the right, leads to energy release) is preferred, resulting in more $^{13}\textrm{CO}$ production, and consequently in a reduced isotope ratio. \corr{This reaction provides the main $^{13}\textrm{CO}$ production and destruction paths in this region. On the other hand, the destruction of $^{12}\textrm{CO}$ is determined by the competing effects of photodissociation, chemical fractionation and dissociative charge transfer with $\textrm{He}^{+}$, while its production is mainly due to the neutral-neutral reaction of light hydrocarbons with oxygen.} At the highest column densities (orange, $A_{\rmn{V}} > 5\,\textrm{mag}$) the 
\corr{CO chemistry is governed by non-isotope-selective reactions. The gas phase production of both CO isotopes happens through $\textrm{HCO}^{+}$ recombination and desorption from dust grains, while the main destruction channels are the dissociative charge transfer with He$^{+}$ and $\textrm{H}_{3}^{+}$, photodissociation by cosmic-ray induced photons, and freeze-out onto grains.}
As a result, the isotope ratio approaches to the elemental ($^{12}$C/$^{13}$C) ratio.
Due to these processes we expect that the CO isotope ratio varies significantly even within the same GMC. In fact, observational studies report a factor of a few region-by-region variation.  
In the case of the Taurus molecular cloud, the indirect measurements of \citet{Goldsmith2008} and \citet{Pineda2010} find isotope ratios between 30 and the canonical value of 69, suggesting $^{13}\textrm{CO}$ enrichment. 

The direct determination of the $^{12}\textrm{CO}/^{13}\textrm{CO}$ ratio is usually difficult and restricted to a certain column density range. For instance, the ultraviolet and millimetre-wavelength absorption measurements, such as presented by \citet{Sheffer2007}, \citet{Sonnentrucker2007} and \citet{LisztLucas1998}, require suitable galactic or extragalactic background sources and CO column densities, which fall into the optically thin, diffuse regime ($N(^{12}\textrm{CO}) < \textrm{few} \times 10^{16}\,\textrm{cm}^{-2}$). 
Observations of millimetre-wavelength emission from GMCs \citep[e.g.][]{Pineda2008,Goldsmith2008,Pineda2010}, however, usually trace the higher CO column density regions, where the isotope ratio-column density correlation is not constrained by observations, and therefore as a ``best guess'' a uniform isotope ratio is adopted.

The inverse problem emerges when $^{13}\textrm{CO}$ emission is inferred from (magneto-)hydrodynamical simulations. 
The computational cost of the chemical modelling scales with the cube of the number of species considered \citep{GloverClark2012b}. Even when only 14 self-consistently calculated (i.e. not described by conservation laws), non-equilibrium species are included in the network, the chemistry will often be the dominating factor in terms of computational cost, taking up to 90 per cent of the total computational time \citep{Glover2010,GloverClark2012b}. For the practical reason of cost efficiency usually only the most common isotope, $^{12}\textrm{CO}$ is included in the chemical networks. When observable quantities, like emission from rarer CO isotopes are inferred from such simulations \citep[e.g.][]{Beaumont2013} the canonical isotope ratio is adopted and assumed to be constant through the whole simulation domain.

In this paper we investigate the effect of \emph{selective photodissociation} and \emph{chemical fractionation} on the $^{12}\textrm{CO}/^{13}\textrm{CO}$ isotopic ratio in different environments and for different cloud properties, using turbulent hydrodynamical simulations that include a self-consistent chemical and cooling model and an approximate treatment of the attenuation of the interstellar radiation field (ISRF). One of our aims is to test and improve the frequently used assumption of uniform $^{12}\textrm{CO}/^{13}\textrm{CO}$ ratio in the context of inferring $^{13}\textrm{CO}$ emission from simulations which neglect isotopic chemistry. We also provide a prescription for deriving the isotope ratio from the $^{13}\textrm{CO}$ column density (e.g. calculated from observations).

In section~\ref{sec:sim} we describe the numerical setup and the initial conditions of our simulations. Section~\ref{sec:ColDenRatio} discusses the correlation between various (total, $^{12}\textrm{CO}$ and $^{13}\textrm{CO}$) column densities and the isotope ratio for different cloud conditions. We also propose a formula for inferring $^{13}\textrm{CO}$ column/number densities from $^{12}\textrm{CO}$ in simulations neglecting fractionation chemistry, and one for calculating the isotope ratio from observations of $^{13}\textrm{CO}$.
Then in section~\ref{sec:EmissionMaps}, we post-process the simulations with line radiative transfer to quantitatively compare the emergent $^{13}\textrm{CO}$ line profiles and emission maps in case of self-consistently calculated, column density dependently inferred and uniform isotope ratios. 
Section~\ref{sec:com} compares our results to previous theoretical works and to observations of the $^{12}\textrm{CO}/^{13}\textrm{CO}$ column density ratio. We summarize the results and draw our final conclusions in section~\ref{sec:Sum}.

\begin{table}
\caption{Model parameters and used snapshots\label{table1}}
\label{tab1}
\begin{center}
\resizebox{8.4cm}{!} {
\begin{tabular}{cccccc}
 \hline
  Model & $n_{0}$ [$\rmn{cm}^{-3}$] & Metallicity [$Z_{\odot}$]& ISRF [$G_{0}$]& Time [Myr] \\
  \hline 
  a & 300 & 0.3 & 1 & 2.046   \\
  b & 300 & 0.6 & 1 & 1.930   \\
  c & 300 & 1   & 0.1 & 2.124 \\
  d & 300 & 1   & 1 & 2.150   \\
  e & 300 & 1   & 10 & 2.022  \\
  f & 1000 & 1  & 1 & 0.973   \\
  \hline
  g & 300 & 1   & 1 & 2.150   \\
  \hline
\end{tabular}
}
\end{center}
Summary of model parameters. Each model cloud has $10^{4}M_{\odot}$ and an SPH mass resolution of 0.5$M_{\odot}$. In each case the analysed snapshots are \corr{chosen to represent the molecular clouds preceding star formation.} $Z_{\odot}$ and $G_{0}$ refer to the solar metallicity and Draine radiation field strength (1.7 in units of the \citet{Habing1968} field) respectively. \corr{Models from a) to f) have fully molecular initial conditions while model g) is calculated with atomic initial composition.}
\end{table}

\section{Simulations} \label{sec:sim}
We use a modified version of the smoothed particle hydrodynamics (SPH) code {\sc gadget-2}\footnote{\protect\url{http://www.mpa-garching.mpg.de/gadget/}}, described by \citet{Springel2005}. The modifications include a sink particle algorithm \corr{to model the formation of individual stars} \citep{Bate1995,Jappsen2005,Federrath2010,GloverClark2012a}, a simplified model of the gas \corr{phase} chemistry with radiative heating and cooling \citep{GloverClark2012b} and an approximate treatment of the attenuation of the ISRF \citep{Clark2012a}.
Stellar feedback from the formed sink particles is not included, an effect which could influence the CO isotope ratio in high density regions significantly. Therefore, we restrict our analysis and discussion to cloud properties before sink particle formation (i.e. to molecular clouds in an early stage of evolution).

\subsection{Chemistry} \label{sec:chem}
We adopt the chemical network of \citet{NelsonLanger99} (hereafter NL99) supplemented with the hydrogen chemistry of \citet{GloverMacLow2007}. The network is designed to follow CO formation and destruction over a wide density range in molecular clouds. It takes multiple CO formation pathways into account: a formation channel involving the composite CH$_{\textrm{x}}$ (CH and CH$_{\textrm{2}}$) species, another involving the composite OH$_{\textrm{x}}$ (OH, H$_{\textrm{2}}$O and O$_{\textrm{2}}$) species and a third route via the dissociative recombination of HCO$^{+}$. The destruction of CO could happen due to photodissociation, dissociative charge transfer with He$^{+}$ or through proton transfer from H$_{3}^{+}$ 
resulting in the conversion of CO to HCO$^{+}$. See Table~1 in \citet{GloverClark2012b} for the full list of reactions and the Appendix B of \citet{Glover2010} for the adopted reaction rate coefficients. 

Initially the NL99 network does not account for the $^{13}$C, $^{13}\textrm{C}^{+}$, $^{13}\textrm{CO}$ and $\textrm{H}^{13}\textrm{CO}^{+}$ isotopes. Therefore we add these species and the corresponding reactions, which are non-isotope-selective, from the original network. We do not apply rescaling on the reaction rate coefficients of these reactions. In addition, to allow conversion of $^{12}\textrm{CO}$ to $^{13}\textrm{CO}$, we implemented the $^{13}\textrm{CO}$ fractionation reaction (equation~\ref{eq:fracreact}, with the left to right rate coefficient of $r_{\rmn{frac,CO,lr}} = 2\times10^{-10} \rmn{cm}^{3}~\rmn{s}^{-1}$ and the temperature dependent right to left rate coefficient of $r_{\rmn{frac,CO,rl}} = 2\times10^{-10} \exp(-35~\rmn{K} / T_{\rmn{gas}})~\rmn{cm}^{3}~\rmn{s}^{-1}$. Although, \citet{SmithAdams1980} suggests an order of magnitude higher rate coefficient, \citet{Sheffer2007} finds that the smaller value of \citet{Watson1976} is more consistent with the observations, therefore we adopt the latter. \corr{The free space $^{12}\textrm{CO}$ and $^{13}\textrm{CO}$ photodissociation rate coefficients in the absence of absorbing material are taken to be equal with the value of $r_{\rmn{pd,CO,thin}} = 2.6\times10^{-10} \rmn{s}^{-1}$ \citep{Visser2009} for the interstellar radiation field strength equal to $1\times\textrm{G}_{\rmn{0}}$}.
\corr{In addition to the reactions included in the \citet{NelsonLanger99} chemical network, we also include the cosmic-ray induced photodissociation of CO \citep{PrasadTarafdar1983,Gredel1987} and the cosmic-ray induced photoionization of C. The reaction rate of the former (i.e. not the rate coefficient) is taken to be $R_{\textrm{CRP,pd,CO}} = 0.21 \times x_{\textrm{H}_{\textrm{2}}} \times \sqrt{T_{\textrm{gas}} x_{\textrm{CO}}}$ and the rate coefficient of the latter is $r_{\textrm{CRP,pi,C}} = 2800 \times \zeta_{\textrm{H}}$ \citep{Maloney1996}. The cosmic ray ionization rate of H is denoted by $\zeta_{\rmn{H}}$ and $x_{\textrm{H}_{\textrm{2}}}$ and $x_{\textrm{CO}}$ are the fractional abundances of the corresponding molecules.}

\corr{In cold and dense molecular clouds the freeze-out onto dust grains becomes the most effective process for removing CO from the gas phase \citep{Bacmann2002,Tafalla2004}. The chemical network presented here does not include this process. This simplification results in the overestimation of gas phase CO abundance, and consequently the CO emissivity in the high density regions. We argue, however, that due to optical depth effects and a relatively low velocity dispersion in the simulated clouds, this probably has a small impact on the CO emission. The freeze-out becomes important only above $A_{\textrm{v}} \approx 6\,\textrm{mag}$ and by this depth, even the $^{13}\textrm{CO}$ emission becomes optically thick, obscuring the freeze-out regions. In any case, the main focus of this work is to constrain the $^{12}\textrm{CO}/^{13}\textrm{CO}$ ratio, which should not be affected by the non-isotope-selective freeze-out process.}

\subsection{Attenuation of the ISRF} \label{sec:isrf}
As we describe in the introduction, the differential shielding of $^{12}\textrm{CO}$ and $^{13}\textrm{CO}$ \corr{is expected to play a role} in determining the isotope ratio. To account for this effect the column densities -- from the centre of a given SPH particle to the outer surface of the cloud -- of the critical species (H$_{2}$, dust, $^{12}\textrm{CO}$ and $^{13}\textrm{CO}$) need to be calculated. We use the TreeCol method presented by \citet{Clark2012a}. In short, TreeCol is a cost efficient algorithm to calculate total and species specific column densities while ``walking'' the gravitational tree structure (used in {\sc gadget-2} to compute the gravitational interaction of far-away particles). It constructs a {\sc healpix} \citep{Gorski2005} sphere with 48 equal-area pixels for each SPH particle and accumulates the contribution of the line of sight nodes for the considered pixel. We trace the total (H nuclei), $\rmn{H}_{2}$, $^{12}\rmn{CO}$ and $^{13}\rmn{CO}$ column densities. The visual extinction ($A_{\rmn{V}}$) due to the dust is calculated using the formula \citep{Bohlin1978,DraineBertoldi1996}, 
\begin{equation}
A_{\rmn{V}} = \frac{N_{\textrm{tot}}}{1.8699\times10^{21} \rmn{cm}^{2}} \times f_{\textrm{dg}},
\end{equation}
where $N_{\textrm{tot}}$ is the total H nuclei column density and $f_{\textrm{dg}} = Z / Z_{\odot}$ is the factor correcting for the simulation metallicity.

We model the attenuation of the ISRF by multiplying the optically thin photodissociation rates with shielding factors depending on the column density and visual extinction. The H$_{2}$ photodissociation rate is attenuated due to dust absorption and H$_{2}$ self-shielding. The dust shielding factor can be calculated in the plane-parallel 
approximation by 
\begin{equation}
\Theta_{\rmn{dust}} = \exp(-\gamma A_{\rmn{V}}) 
\label{eq:dustshield}
\end{equation}
with $\gamma=3.74$ \citep{DraineBertoldi1996}. The self-shielding factor depends on the H$_{2}$ column density and is calculated according to equation~(37) in \citet{DraineBertoldi1996}. In the case of $^{12}\textrm{CO}$ and $^{13}\textrm{CO}$, the shielding is due to dust absorption, the H$_{2}$ Lyman-Werner lines and CO self-shielding. The shielding factor due to dust absorption for both isotopic species is given by equation~(\ref{eq:dustshield}) with \corr{$\gamma=3.53$. The tabulated CO shielding by $\textrm{H}_{2}$ and self-shielding factors are adopted from \citet{Visser2009}.} We used the same relation between the column density and the CO self-shielding factor for both CO isotopes when calculating the self-shielding, but with the corresponding isotope column density.

We refer to section 2.2 in \citet{Glover2010} for a more detailed description of the adopted treatment of photochemistry.

\subsection{Thermal model} \label{sec:thermal}
\corr{We calculate the thermal balance and temperatures of gas and dust self-consistently, taking a number of cooling and heating processes into account. The temperature structure of the cloud has a large impact on the chemical fractionation (due to the temperature barrier for the right-to-left path), a moderate effect on the selective photodissociation (more strongly isotope selective in colder gas; discussed in detail in \citet{Visser2009}, but not considered here) and significant influence on CO excitation, therefore a realistic thermal model is necessary.}

\begin{figure}
\includegraphics[scale=0.28]{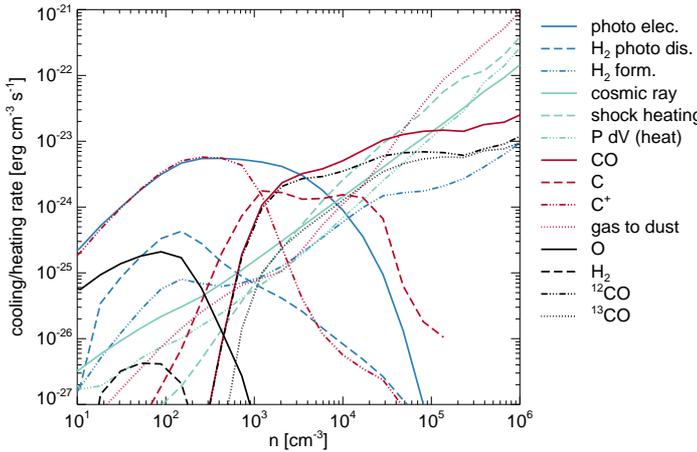}
\caption{\corr{The median cooling and heating rates as a function of hydrogen nuclei number density for the fiducial model (d). The cooling rates of C-bearing molecules are calculated separately for the $^{12}\textrm{C}$ and $^{13}\textrm{C}$ isotopologues but their combined rates are shown here. The exception is CO, for which we also plot the isotopic contributions separately. For the detailed description of the cooling and heating processes see \citet{GloverClark2012a}.}}
\label{fig:heatcool}
\end{figure}

\corr{The adopted thermal model -- with the complete list of heating and cooling processes and rates -- is presented in section 3.2.4 of \citet{GloverClark2012a}. Fig.~\ref{fig:heatcool} summarizes the contributions of thermal processes to the thermal balance in our simulations. The dominant heating processes are the photoelectric, shock, cosmic ray, and P dV (expansion and contraction) heating. The major coolants at low and intermediate densities are $\textrm{C}^{+}$ and CO, while at high density cooling is dominated by the dust. We consider the $\textrm{C}^{+}$ and CO isotopes separately when calculating the cooling rates. Thanks to its lower optical depth, $^{13}\textrm{CO}$ might became as effective coolant as $^{12}\textrm{CO}$ at densities higher than $10^{4}\,\textrm{cm}^{-3}$. However, the effect of this on the thermal balance is negligible, since at these densities, dust cooling is already the dominant process.}

\subsection{Initial conditions} \label{sec:initcond}
Our basic initial setup is identical to \citet{GloverClark2012a}. We start the simulations with a uniform density sphere with $10^4M_{\odot}$ total mass. The initial volume density of the sphere is set to 300 or 1000 cm$^{-3}$, resulting in an approximate cloud radius of 6 pc or 4 pc, respectively. \corr{The initial velocities of the SPH particles are chosen so that the initial velocity field has a steep power spectrum with $\rmn{P}(k) \propto k^{-4}$, using the ``cloud-in-cell'' scheme \citep{HockneyEastwood1988,MacLow1998}. The velocity field is scaled such that the total kinetic energy equals to the gravitational potential energy, corresponding to 3D root-mean-square (rms) velocities ($\sigma_{\textrm{rms,3D}}$) of $2.81\,\textrm{km s}^{-1}$ and $3.43\,\textrm{km s}^{-1}$ for the different initial densities.
We do not apply turbulent driving during the simulations and the turbulence is allowed to dissipate freely through shocks and numerical viscosity. Consequently, the overall rms velocity of the cloud decreases with time. In the first ${\sim}0.5$ Myr the rms velocity stagnates; afterwards, it decreases with time as $\sigma_{\textrm{rms,3D}}(t) \propto t^{-0.24}$. By the time of the analysed snapshots it reaches the values of 2 and $2.6\,\textrm{km s}^{-1}$}.
The initial gas and dust temperature are uniform at 20 K and 15 K, respectively.

In case of the solar metallicity ($Z_{\odot}$) runs, the adopted initial abundances of $^{12}$C, $^{13}$C and O relative to hydrogen nuclei are $x_{\rmn{^{12}C}}=1.4\times10^{-4}$, $x_{\rmn{^{13}C}}=2.3\times10^{-6}$ and $x_{\rmn{O}}=3.2\times10^{-4}$ respectively \citep[][]{Sembach2000}. \corr{In the case of ``fully molecular'' initial composition (models a to f), all hydrogen atoms are in $\textrm{H}_{2}$ form, while when ``atomic'' initial conditions are adopted (model g, see discussion in Appendix~\ref{sec:initchemcomp}), then all hydrogen is atomic and neutral. In both cases we assume that all carbon is in ionized form.} The helium is neutral and its fractional abundance is \corr{0.079, equivalent to 24 per cent mass fraction,} in all simulations. The total abundance of low ionization potential metals (Na, Mg, etc.) is $x_{\textrm{M}}=1\times10^{-7}$. These are initially assumed to be fully ionized. The electron abundance of the cloud is set to give an overall neutral medium. To investigate the metallicity dependence of our results, we also perform simulations with scaled initial abundances geared towards the SMC ($0.3 \times Z_{\odot}$) and the LMC ($0.6 \times Z_{\odot}$). In each run the carbon isotopic abundance ratio is 60, a value consistent with the frequently adopted \corr{measurements \citep{LucasLiszt1998} in the solar vicinity}. Besides the abundances we also scale the dust-to-gas ratio with a factor, $f_{\textrm{dg}} = Z / Z_{\odot}$. In the solar metallicity case the dust-to-gas ratio is taken to be 0.01.

We assume that the cloud is illuminated by an isotropic, standard radiation field, described by \citet{Draine1978} in the UV and by \citet{Black1994} at longer wavelengths. The default field strength is $G_{0} = 1.7$ in units of the \citet{Habing1968} field, corresponding to $2.7\times10^{-3}~\rmn{erg}~\rmn{cm}^{-2}~\rmn{s}^{-1}$ integrated flux in the 91.2--240 nm wavelength range. To test various galactic environments we scaled the radiation field strength between $0.1 \times G_0$ and $10 \times G_0$ with the fiducial value of $1 \times G_0$.

\corr{In the most shielded regions of GMCs, several important CO destruction pathways are activated by the deep penetrating cosmic ray particles, which also provide an important gas heating mechanism there.} The adopted cosmic ray ionization rate of atomic hydrogen is $\zeta_{\rmn{H}}=10^{-17}$s$^{-1}$. The ones for $\rmn{H}_{2}$ and atomic He are $2\times\zeta_{\rmn{H}}$ and $1.09\times\zeta_{\rmn{H}}$. respectively.

We performed 6+1 hydrodynamic cloud simulations in total. The parameters explored are summarized in Table~\ref{tab1}. In the analysis that follows in section~\ref{sec:ColDenRatio} we take simulation d) as the fiducial model.

\section{The \texorpdfstring{$^{12}\textrm{CO}/^{13}\textrm{CO}$}{12CO/13CO} column density ratio} \label{sec:ColDenRatio}

\begin{figure*}
\includegraphics[scale=0.52]{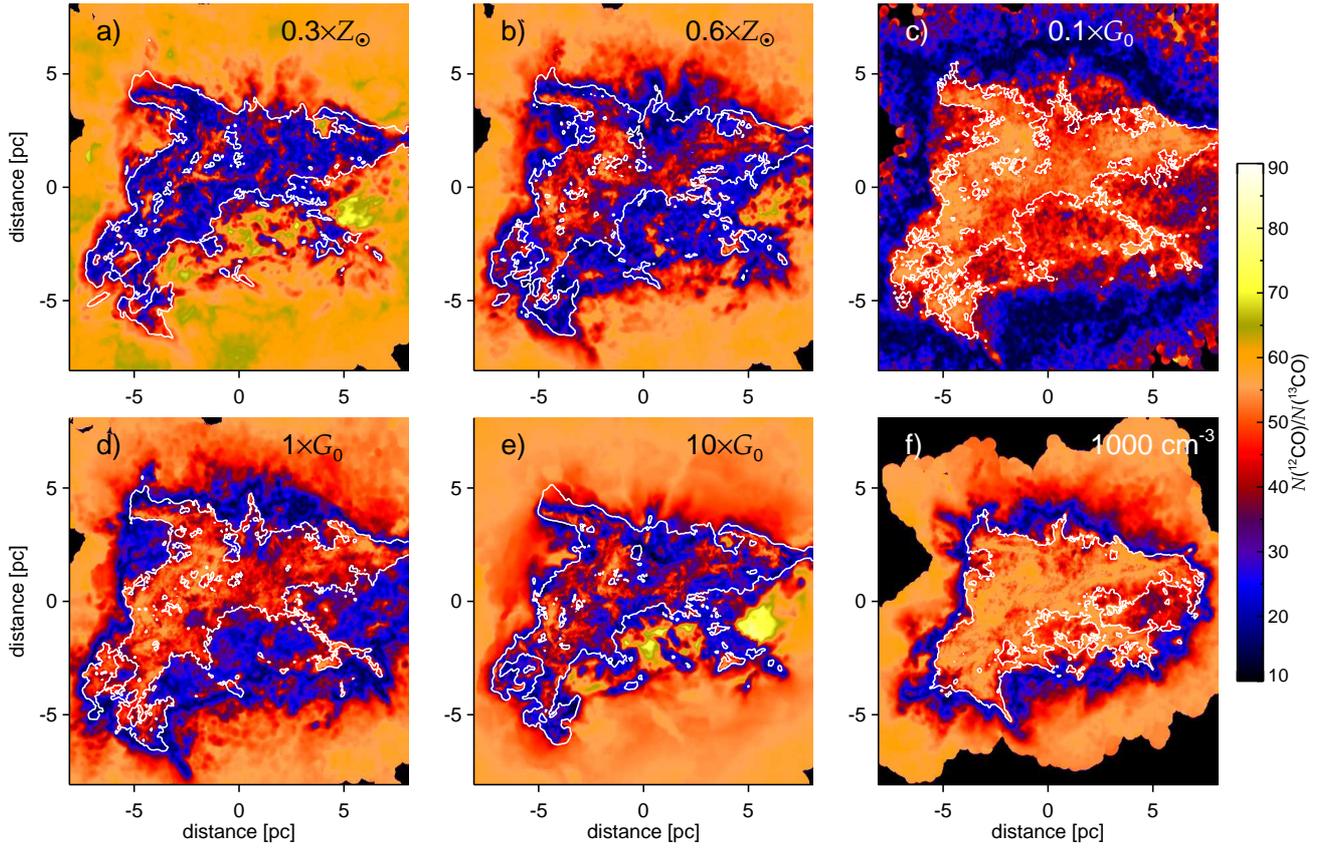}
\caption{$N(^{12}\textrm{CO})/N(^{13}\textrm{CO})$ column density ratio maps for varying ISRFs, metallicities and initial volume densities. The white contour represents the $5\times10^{21}$ cm$^{-2}$ level in total column density. The same colour scale applies for all the ratio maps. Yellow and blue colors mark regions dominated by the effects of selective photodissociation and chemical fractionation, respectively, while orange marks regions in which the CO is not significantly fractionated. Lines of sight which do not intersect any SPH particles are marked with black.}
\label{fig:isoratio}
\end{figure*}

In this section we analyse the $N(^{12}\textrm{CO})/N(^{13}\textrm{CO})$ isotopic column density ratio resulting from our simulations. We look for correlations between the isotope ratio and the total, $^{12}\textrm{CO}$ and $^{13}\textrm{CO}$ column densities. Hereafter, if not indicated otherwise, the column densities are meant as the number of atoms/molecules of a species or the gas mass in a given line of sight, per $\rmn{cm}^{2}$, integrated from the ``observer'' to infinity.

We use the grid interpolated quantities (see Appendix~\ref{sec:grid}). The column density maps are calculated by integrating the volume density along the $z$ direction of the grid. Although the apparent shape of the molecular cloud does depend on the viewing direction, the quantities and relationships discussed below are independent of the choice of viewing direction. Fig.~\ref{fig:isoratio} shows the ratio of $^{12}\textrm{CO}$ and $^{13}\textrm{CO}$ column density maps for different radiation fields, metallicities and initial volume densities. The white contour line indicates the total column density level of $5\times10^{21}\,\textrm{cm}^{-2}$. The shape of this contour line does not change substantially with varying radiation field strength or metallicity, indicating that the overall density structure is not affected significantly within the studied parameter range (see also Fig.~\ref{fig:pdfsad}).

The simulated ratio maps are in \corr{general agreement} with the picture of the CO isotopic chemistry described in the introduction. In the outer parts of the clouds, the total column densities are low and the shielding of neither isotopic species is effective. \corr{Deeper in the cloud (see model a and e), the selective photodissociation of $^{13}\textrm{CO}$ might dominate, increasing the $^{12}\textrm{CO}/^{13}\textrm{CO}$ ratio up to ${\sim}75$ (yellow region). Further in, where most of the cloud mass resides, the fractionation reaction takes over and significantly decreases the ratio to ${\sim}25$ and below (blue region)}. At the core of the cloud, the radiation field photodissociates CO or ionizes C with a very low rate, and so neither of the isotope-selective processes are effective. 
In these regions the $N(^{12}\textrm{CO})/N(^{13}\textrm{CO})$ ratio increases again and approaches the initially set $^{12}\textrm{C}/^{13}\textrm{C}$ ratio (marked by the orange color). Qualitatively all simulations follow this scheme. However, the quantitative details -- e.g the transitional column densities  -- differ considerably from simulation to simulation.

\begin{figure*}
\includegraphics[scale=0.535]{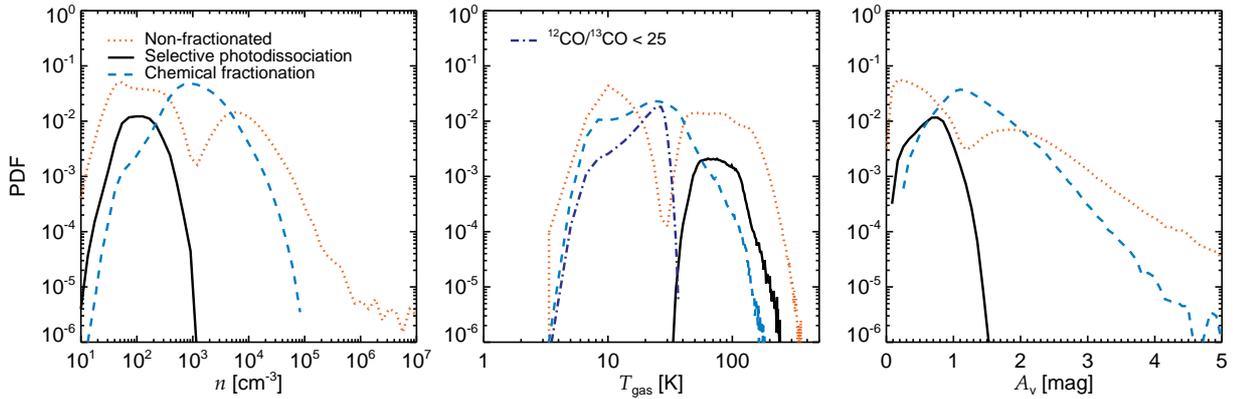}
\caption{\corr{The number density, gas temperature and mean visual extinction PDFs of the non-fractionated ($50 < {^{12}\textrm{CO}}/{^{13}\textrm{CO}} < 65$), chemical fractionation dominated ($^{12}\textrm{CO}/^{13}\textrm{CO} \leq 50$) and isotope selective photodissociation dominated ($^{12}\textrm{CO}/^{13}\textrm{CO} \geq 65$) zones, weighted by the mass. Simulation e) is shown here, but all the other models qualitatively follow the same distributions.}}
\label{fig:pdf}
\end{figure*}

\corr{To investigate the typical physical conditions of the three zones (non-fractionated or dominated by the effects of selective photodissociation or chemical fractionation) we plot their number density, gas temperature and mean\footnote{\corr{Average over the 48 directions of the {\sc healpix} sky of each SPH particle.}} visual extinction probability density functions (PDFs) in Fig.~\ref{fig:pdf}. The non-fractionated gas shows bimodal distributions in all three quantities. At low number densities, CO is poorly shielded from the effects of the ISRF and both CO isotopes are photodissociated at approximately the same rate. In this regime, photodissociation is non-selective and does not significantly alter the $^{12}\textrm{CO}/^{13}\textrm{CO}$ ratio. At high number densities ($n > 10^{4} \: {\rm cm^{-3}}$), the ISRF is strongly attenuated and almost all of the available carbon is locked up in CO. In this regime, the $^{12}\textrm{CO}/^{13}\textrm{CO}$ ratio is necessarily close to the elemental $^{12}\textrm{C}/^{13}\textrm{C}$ ratio. The effects of isotope-selective photodissociation peak at number densities around $100\,\textrm{cm}^{-3}$, high temperatures and moderate visual extinctions (between 0.5 and 1 $\textrm{mag}$). However, even in this parameter regime, only a small fraction of the gas has an isotope ratio 65 or higher. This high ratio gas resides in a thin layer around the bulk of the cloud. Normally this thin layer is not visible in the column density maps, since such maps trace a mass-weighted ratio along a given line of sight. In certain cases (see models a and e), high ratios could still be found, if most of the gas in the line of sight is relatively dilute (a few hundred particles per $\textrm{cm}^{3}$) and moderately shielded (about $0.5\,\textrm{mag}$ visual extinction).
Chemical fractionation dominates in gas which has a characteristic number density between $10^{3}$ and $10^{4}\,\textrm{cm}^{-3}$ and mean visual extinction around unity. The chemically fractionated gas shows a wide temperature range. However, the temperature distribution of the highly fractionated gas (ratio lower than 25, dotted dashed line on the middle panel of Fig.~\ref{fig:pdf}) has an abrupt cut-off around 30 K.}

\corr{The dependence of the ${^{12}\textrm{CO}}/{^{13}\textrm{CO}}$ ratio on the physical conditions in the cloud that we find from our study is in a good qualitative agreement with the large parameter study of photon dominated region (PDR) models in \citet{Roellig2013}. We conclude that effect of the isotope selective photodissociation is negligible when column density isotope ratios are considered. Hereafter, we mainly focus on the effect of chemical fractionation.}

\begin{figure}
\begin{center}
\includegraphics[scale=0.6]{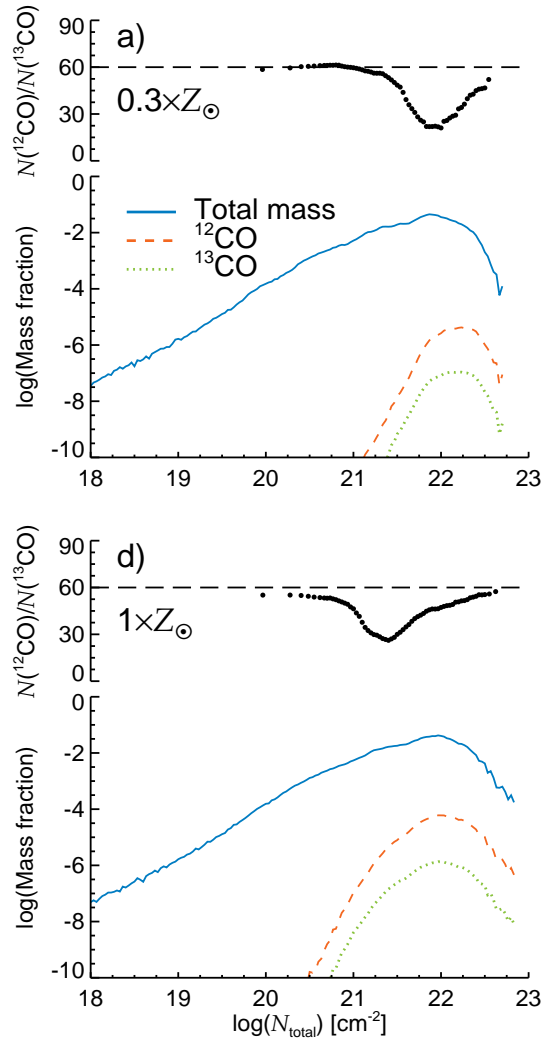} 
\end{center}
\caption{Total, $^{12}\textrm{CO}$ and $^{13}\textrm{CO}$ mass fraction distributions and the $N(^{12}\textrm{CO})/N(^{13}\textrm{CO})$ ratio as a function of the total column density for low metallicity (a) and the fiducial (d) simulations. The total column density distributions are roughly identical, while CO forms at a factor of 3 higher column densities in the case of a). The total CO mass is also reduced by about an order of magnitude. The dip in the isotope ratio curve is shifted by almost a order of magnitude to higher total column densities. Note that the column density ratio is not expected to be equivalent to the ratio of the $^{12}\textrm{CO}$ and $^{13}\textrm{CO}$ mass distributions, as in the latter case the area associated with a column density value is also taken into account.}
\label{fig:pdfsad}
\end{figure}

\subsection{Correlation with the total column density} \label{sec:ratiovstotcol}

The mass-weighted PDFs of the total, the $^{12}\textrm{CO}$ and the $^{13}\textrm{CO}$ column densities for simulations a) and d) are presented in Fig.~\ref{fig:pdfsad}. The figure shows the mass fraction of the cloud at a given column density. The CO isotope column densities are weighted with the total mass, therefore the corresponding curves show the cloud mass fraction locked in CO isotopes as a function of the total column density.

The overall density distributions in these simulations are roughly identical. The CO isotope distributions, however, show large differences: due to less effective shielding, in simulation a) both CO isotopes form at a factor of a few higher total column densities and the total $^{12}\textrm{CO}$ mass (the integral of the red curve multiplied by the total cloud mass) is also a factor of \corr{17 lower, 0.6 M$_{\odot}$ and 10.2 M$_{\odot}$} for simulation a) and d) respectively. The total $^{13}\textrm{CO}$ mass is reduced by a similar factor. Consequently, depending on cloud properties and environment, CO isotopes trace different total column densities and cloud mass fractions.
Also note the relatively large fraction of the so called CO-dark gas, i.e. mass not traced at all by CO \citep{Wolfire2010,Smith2014}.

The upper sub-panels in Fig.~\ref{fig:pdfsad} show the mean $N(^{12}\textrm{CO})/N(^{13}\textrm{CO})$ ratio as a function of the total column density. The description of the algorithm that we used to produce these curves is given in section~\ref{sec:fitting}. The gradients of the curves represent the transitions between regions dominated \corr{by photodissociation}, chemical fractionation and strong attenuation. The total column densities corresponding to these transitions are strongly dependent on the strength of the ISRF and the metallicity of the cloud. In case of simulation d) the $5\times10^{21}$ cm$^{-2}$ total column density contour (see Fig.~\ref{fig:isoratio}) approximately indicates the transition between the regions dominated by chemical fractionation and strong attenuation. In the case of simulation a), the reduced metallicity decreases both the CO column density and the extinction due to dust that a parcel of gas in the cloud ``sees'', resulting in less effective attenuation of the radiation field. The transition between the chemical fractionation-dominated and the attenuation-dominated regions is shifted to a total column density that is an order of magnitude larger. In this case, the $5\times10^{21}$ cm$^{-2}$ contour line traces the transition between the regions dominated by \corr{non-selective} photodissociation and chemical fractionation. Simulations with reduced metallicity or increased ISRF (b and e) behave similarly to model a), while simulation f) has transitional column densities similar to model d). \corr{In case of c), due to the weak radiation field, a substantial amount of CO is able to form even at low visual extinctions. This results in an extended chemical fractionation dominated region at very low total column densities (i.e. most likely below any reasonable detection limit) and a very strong shielding, and hence close to elemental ratios in the (observable) inner regions.}

Regardless of the model dependent transitional total column densities, the isotope ratio seems to be consistent within a few per cent in the corresponding regions of the cloud for each simulation. \corr{In the region dominated by non-selective photodissociation the ratio is close to the initially set $^{12}$C/$^{13}$C ratio of 60. In the region of chemical fractionation, the ratio drops to ${\sim}20$. Finally, in the most shielded regions the CO isotope ratio approaches 60 again}. Note, however, that the results from the outermost regions of the cloud are -- due to large smoothing lengths and low SPH resolution -- somewhat uncertain. These regions are usually also undetectable in CO emission, because of their very low column densities.

Fig.~\ref{fig:pdfsad} also shows that the region of $^{13}\textrm{CO}$ enhancement coincides with the highest CO mass fraction in case of simulation a) and with lower mass fractions in case of d). This indicates a higher importance of isotope-selective reactions when the metallicity is low.

We conclude that the total column density and the isotope ratio correlate, but the correlation strongly depends on the cloud properties and environment.

\subsection{Correlation with the \texorpdfstring{$^{12}\textrm{CO}$}{12CO} column density} \label{sec:ratiovsn12co}

\begin{figure*}
\includegraphics[scale=0.52]{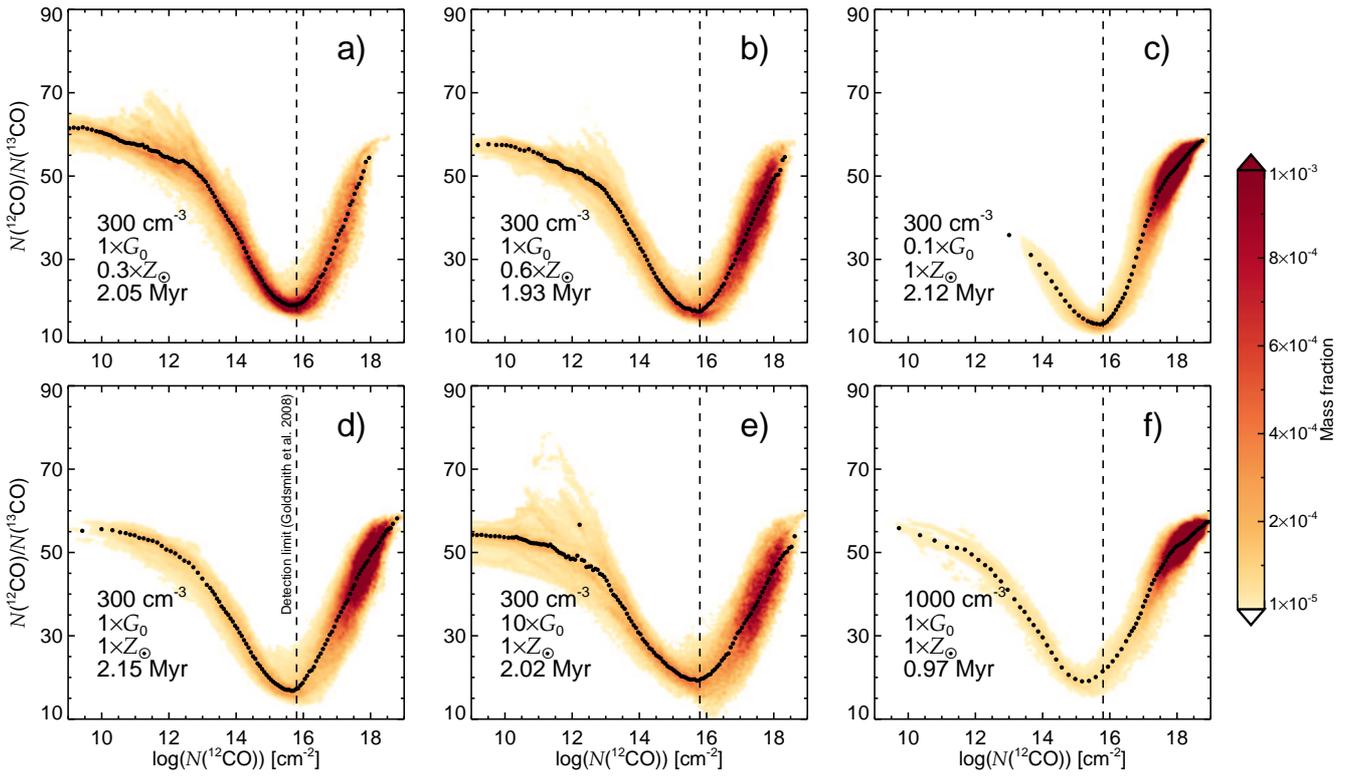}
\caption{CO isotope column density ratio as a function of the $^{12}\textrm{CO}$ column density, calculated using the ``real'' column densities (i.e. directly from the hydrodynamic simulation). Model parameters are indicated in the legend, the vertical line shows the approximate detection limit achieved by \citet{Goldsmith2008} in the case of the Taurus molecular cloud region. The colour indicates the mass fraction of the cloud with the specified $^{12}\textrm{CO}$ column density and isotope ratio, which depends significantly on the model parameters. 
At low column densities, isotope-selective photodissociation might increase the ratio up to 70 in some cases. At higher column densities, the fractionation reaction takes over, resulting in a pronounced dip. The observations are sensitive to the inner regions of the cloud, where the fractionation reaction becomes less effective as the C$^{+}$ abundance decreases, and the ratio increases to the initial value. The black dotted curves represent the fitted relation between the quantities (see Section~\ref{sec:fitting}).}
\label{fig:2d_isorat_12}
\end{figure*}

We find a tighter correlation between the $^{12}\textrm{CO}$ column density and the CO isotope ratio. Fig.~\ref{fig:2d_isorat_12} shows the mass-weighted, two dimensional probability density distribution of $N(^{12}\textrm{CO})$ and the isotope ratio for the 6 simulations (from a to f). The colours indicate the mass fraction of the cloud with a given parameter combination. Consistent with Fig.~\ref{fig:pdfsad}, most of the cloud mass lies at different $^{12}\textrm{CO}$ column densities, depending on the model parameters. If the metallicity is low or the radiation field is strong, then CO forms at, and therefore traces, higher total column densities.

\begin{figure*}
\includegraphics[scale=0.535]{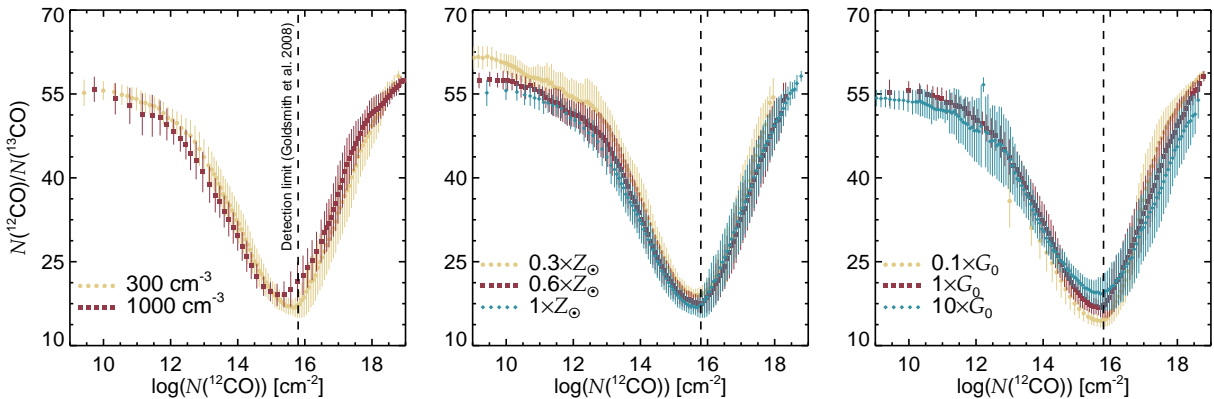}
\caption{CO isotope column density ratio as a function of the $^{12}\textrm{CO}$ column density derived by the procedure described in section~\ref{sec:ratiovsn12co}. From left to right the panels compare the effect of varying the initial density, metallicity and radiation field. We emphasize that the main features of the curves only weakly depend on the physical parameters. Model parameters not indicated in the legend are kept at the fiducial value.}
\label{fig:isorat_12curve}
\end{figure*}

The most remarkable feature of Fig.~\ref{fig:2d_isorat_12} is that the location and depth of the dip in the isotope ratio shows only weak parameter dependence. The three dimensional nature ($N(^{12}\textrm{CO})$ vs. isotope ratio vs. mass fraction) of these probability density distribution diagrams, however, makes it hard to directly compare them. Therefore, we project them into 2D space, keeping in mind that we are interested in the relationship of $N(^{12}\textrm{CO})$ and the isotope ratio, and that we are aiming to derive a functional form. We construct curves according the following procedure: Starting from the low $^{12}\textrm{CO}$ column densities we bin $N(^{12}\textrm{CO})$ adaptively into strips. The normalized total mass fraction of a strip is required to be at least 0.002, a value chosen to provide well sampled isotope ratio distribution in the strip. We then collapse the strip in the $N(^{12}\textrm{CO})$ dimension by adding up mass fractions in isotope ratio bins. This gives the mass fraction in a $N(^{12}\textrm{CO})$ strip as the function of isotope ratio. We fit a Gaussian to this curve to determine the mean value and the width of the distribution (i.e. the standard deviation). Finally, we determine an effective $N(^{12}\textrm{CO})$ within the strip, by weighting based on the mass fraction contribution of a $N(^{12}\textrm{CO})$ pixel column to the strip. We repeat this procedure until the upper limit of the $^{12}\textrm{CO}$ column density is reached. The small, filled, black circles in the panels in Fig.~\ref{fig:2d_isorat_12} and the coloured filled symbols in Fig.~\ref{fig:isorat_12curve} show the mean isotope ratios as a function of $^{12}\textrm{CO}$ column density derived this way. Fig.~\ref{fig:2d_isorat_12} demonstrates that the derived mean ratios follow the probability density distribution well, justifying our approach. 

The mean isotope ratio curves show a very good overall agreement in all cases (see Fig.~\ref{fig:isorat_12curve}). However, there is a weak correlation between the curve shape and the physical parameters. A higher initial density ($n_{0}$) might result in higher isotope ratios at high $^{12}\textrm{CO}$ column densities. As the metallicity decreases from $1\,Z_{\odot}$ to $0.3\,Z_{\odot}$ the isotope selective photodossociation seem to increase the $^{12}\textrm{CO}/^{13}\textrm{CO}$ ratio slightly above the elemental at CO column densities below $10^{11} \textrm{cm}^{-2}$ (middle panel). The decreasing radiation field strength seems to decrease the minimum isotope ratio from $19.23\pm2.98$ to $14.40\pm1.17$ (right panel).

The depth of the dip in the isotope ratio curve is expected to depend on the gas temperature (and therefore the heating and cooling processes) of the corresponding cloud regions. The chemical fractionation has an energy barrier for the right to left reaction path (see equation~\ref{eq:fracreact} and section~\ref{sec:chem}), which is approximately 35 K. At temperature much higher than this, the reaction could proceed in both directions with similar rate, resulting in a less enhanced $^{13}\textrm{CO}$ abundance and a isotope ratio closer to the $^{12}$C/$^{13}$C ratio \citep[e.g.][]{Roellig2013}.

We emphasize, however, that these trends are not statistically significant in our simulations, and the model-by-model deviations of the isotope ratio curves are typically comparable to the standard deviations of the ratio. From this point on, we dispense with further investigation of the trends in the mean isotope ratio curves with physical parameters and assume that there is an unequivocal correlation between the $^{12}\textrm{CO}$ column density and the $^{12}\textrm{CO}/^{13}\textrm{CO}$ ratio which is independent of the parameters we vary in the simulations.

\begin{figure}
\includegraphics[scale=0.535]{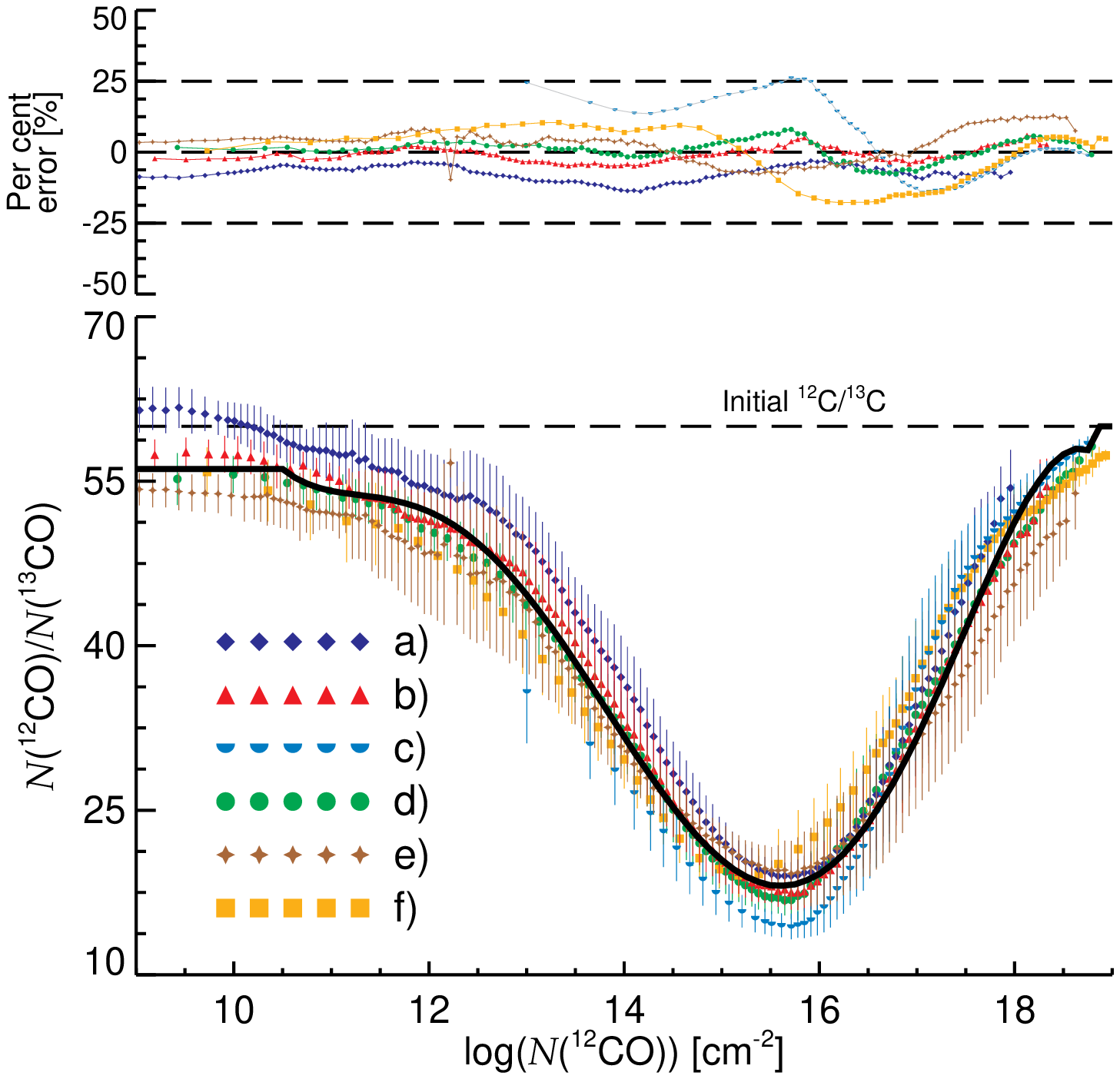}
\caption{Similar to Fig.~\ref{fig:isorat_12curve}. Here we compare the $N(^{12}\textrm{CO})$-isotope ratio curves derived from the simulation (coloured dotted lines) to the adopted fitting formula (black solid line). The upper panel shows the percentage error of the formula when compared to the model data.}
\label{fig:fittingformula}
\end{figure}

\subsection{Fitting formula} \label{sec:fitting}

To derive a functional form for the $N(^{12}\textrm{CO})$-isotope ratio relationship we fit the curves presented in Fig.~\ref{fig:isorat_12curve} individually and together using the non-linear least-squares Marquardt-Levenberg algorithm implemented in {\sc gnuplot}\footnote{\protect\url{http://gnuplot.sourceforge.net/}} \citep{Williams2011}. \corr{The combined data are constructed from all models except models c) and g), due to the low numerical resolution at low column densities in the former and for the sake of consistency in the latter case.}
We fitted 4th, 5th and 6th order polynomial functions taking the standard deviation of each data point into account. The best fit 4th order polynomial over-predicts the ratios for $10^{15}\,\rmn{cm}^{-2}\,<\,N(^{12}\textrm{CO})\,<\,10^{17}\,\rmn{cm}^{-2}$, and under-predicts in every other case. The 6th order polynomial does not provide a significantly better fit than the 5th order polynomial, therefore we chose to use the best fitting 5th order polynomial for the further analysis. The best fitting polynomial coefficients are presented in Table~\ref{bestfit} for the individual and the combined data.

\begin{table*}
\caption{Coefficients of the best fitting polynomials in case of the $^{12}\textrm{CO}$ column density--isotope ratio correlation\label{table3}}
\label{bestfit}
\begin{tabular}{ccccccc}
  \hline
   Model & $a_{0,12}$ & $a_{1,12}$ & $a_{2,12}$ & $a_{3,12}$ & $a_{4,12}$ & $a_{5,12}$        \\
  \hline
a        & 44118.20$\pm$946.40    & -16028.90$\pm$336.30  & 2308.65$\pm$47.51
         & -164.3870$\pm$3.336        &  5.78127$\pm$0.1164          &  -0.080301$\pm$0.00162 \\
b        & 39398.60$\pm$990.50     & -14339.70$\pm$348.80  & 2069.68$\pm$48.80
         & -147.7000$\pm$3.390        &  5.20628$\pm$0.1170       &  -0.072481$\pm$0.00160 \\
c        & -129934.00$\pm$22970.00 & 39353.20$\pm$7218.00  & -4717.00$\pm$904.00 
         & 279.7200$\pm$56.410       &  -8.20654$\pm$1.7540         & 0.095312$\pm$0.02170   \\
d        & 23435.20$\pm$1973.00    & -8696.25$\pm$675.90  & 1277.96$\pm$91.93
         & -92.6067$\pm$6.206        &  3.30431$\pm$0.2079       & -0.046415$\pm$0.00277  \\
e        & 16928.80$\pm$919.10    & -6337.57$\pm$320.20  & 939.74$\pm$44.29
         & -68.6503$\pm$3.041        &  2.46668$\pm$0.1037       & -0.034852$\pm$0.00141  \\
f        & 6002.58$\pm$2009.00   & -2530.75$\pm$701.70  & 416.65$\pm$96.97 
         & -33.2072$\pm$6.631        &  1.28180$\pm$0.2244       & -0.019203$\pm$0.00301  \\
\hline
combined & 23619.40$\pm$1657.00    & -8801.37$\pm$566.20  & 1298.19$\pm$76.76 
         & -94.3795$\pm$5.162        &  3.37743$\pm$0.1723       &  -0.047569$\pm$0.00228 \\
 \hline
\end{tabular}
\end{table*}

\begin{table*}
\caption{Coefficients of the best fitting polynomials in case of the $^{13}\textrm{CO}$ column density--isotope ratio correlation\label{table4}}
\label{bestfit13co}
\begin{tabular}{ccccccc}
  \hline
   Model & $a_{0,13}$ & $a_{1,13}$ & $a_{2,13}$ & $a_{3,13}$ & $a_{4,13}$ & $a_{5,13}$        \\
  \hline
a        & 3430.54$\pm$304.30    & -1389.82$\pm$132.70  & 220.92$\pm$22.84
         & -16.7171$\pm$1.941        &  0.59055$\pm$0.0815          &  -0.007564$\pm$0.00135 \\
b        & 4607.55$\pm$476.30     & -1955.96$\pm$206.00  & 328.11$\pm$35.16
         & -26.7192$\pm$2.961        &  1.05021$\pm$0.1230       & -0.015883$\pm$0.00202 \\
c        & 57181.50$\pm$28280.00 & -22455.80$\pm$9957.00  & 3494.39$\pm$1397.00 
         & -269.0400$\pm$97.540       &  10.24410$\pm$3.3920         & -0.154282$\pm$0.04700   \\
d        & 9050.64$\pm$665.40    & -3887.04$\pm$280.10  & 659.215$\pm$46.52
         & -54.7241$\pm$3.813        &  2.21884$\pm$0.1542       & -0.035134$\pm$0.00247  \\
e        & 4956.36$\pm$272.80    & -2159.25$\pm$117.60  & 372.237$\pm$19.98
         & -31.2920$\pm$1.674        &  1.27851$\pm$0.0691       & -0.020294$\pm$0.00113  \\
f        & 12412.20$\pm$796.60   & -5319.18$\pm$323.70  & 901.857$\pm$51.93 
         & -75.1591$\pm$4.112        &  3.07318$\pm$0.1608       & -0.049292$\pm$0.00249  \\
\hline
combined & 8519.51$\pm$411.00    & -3677.23$\pm$174.70  & 626.981$\pm$29.30 
         & -52.3141$\pm$2.423        &  2.13123$\pm$0.0988       &  -0.033897$\pm$0.00159 \\
\hline
\end{tabular}
\end{table*}

We adopt the best fit coefficients for the combined model with the modifications that the ratio at $^{12}\textrm{CO}$ column densities lower than $3.2 \times 10^{10} \rmn{cm}^{-2}$ is equal to the ratio at $N(^{12}\textrm{CO}) = 3.2 \times 10^{10} \rmn{cm}^{-2}$ and we set the ratio to 60 above an upper limit of $N(^{12}\textrm{CO}) = 6 \times 10^{18} \rmn{cm}^{-2}$. 
The final form of our fitting formula is 
\begin{equation}
r(N_{12}) = \begin{cases} r(N_{12}= 3.2 \times 10^{10})
&\mbox{if } N_{12} < 3.2 \times 10^{10} \\
    a_{0,12} + a_{1,12} \log_{10}(N_{12}) \\ + a_{2,12} \log_{10}(N_{12})^2 \\ + a_{3,12} \log_{10}(N_{12})^3 \\ + a_{4,12} \log_{10}(N_{12})^4  \\ + a_{5,12} \log_{10}(N_{12})^5 & \mbox{if } 3.2 \times 10^{10} \leq N_{12} \leq 6 \times 10^{18} \\                 
    60              & \mbox{if } N_{12} > 6 \times 10^{18}  \end{cases} 
\label{eq:fitfun}
\end{equation}
where $r$ is the $^{12}\textrm{CO}/^{13}\textrm{CO}$ ratio and $N_{12}$ is the $^{12}\textrm{CO}$ column density in units of $\textrm{cm}^{-2}$. To justify this choice we compare the function derived from the combined models to the individual models. The bottom panel of Fig.~\ref{fig:fittingformula} shows the unmodified curves from section~\ref{sec:ratiovsn12co} for all models with the standard deviations represented by the vertical lines. The black line represent the fitting function. The top panel show the per cent error between the data points from the models and the approximate value from the formula, $\Delta = (function - data)/data \times 100$.
The deviation with which the formula reproduces the data tends to be the largest in the fractionation reaction dominated region ($10^{15} \rmn{cm}^{-2} < N(^{12}\rmn{CO}) < 10^{16} \rmn{cm}^{-2}$) going up to about 25 per cent. Note however, the difference on the whole $^{12}\textrm{CO}$ column density range ($\leq15$ per cent) is comparable to the standard deviation (i.e. the thickness) of the combined distribution. 

\subsection{Application for \texorpdfstring{$^{13}\textrm{CO}$}{13CO} observations} \label{sec:13coobs}

\begin{figure*}
\includegraphics[scale=0.52]{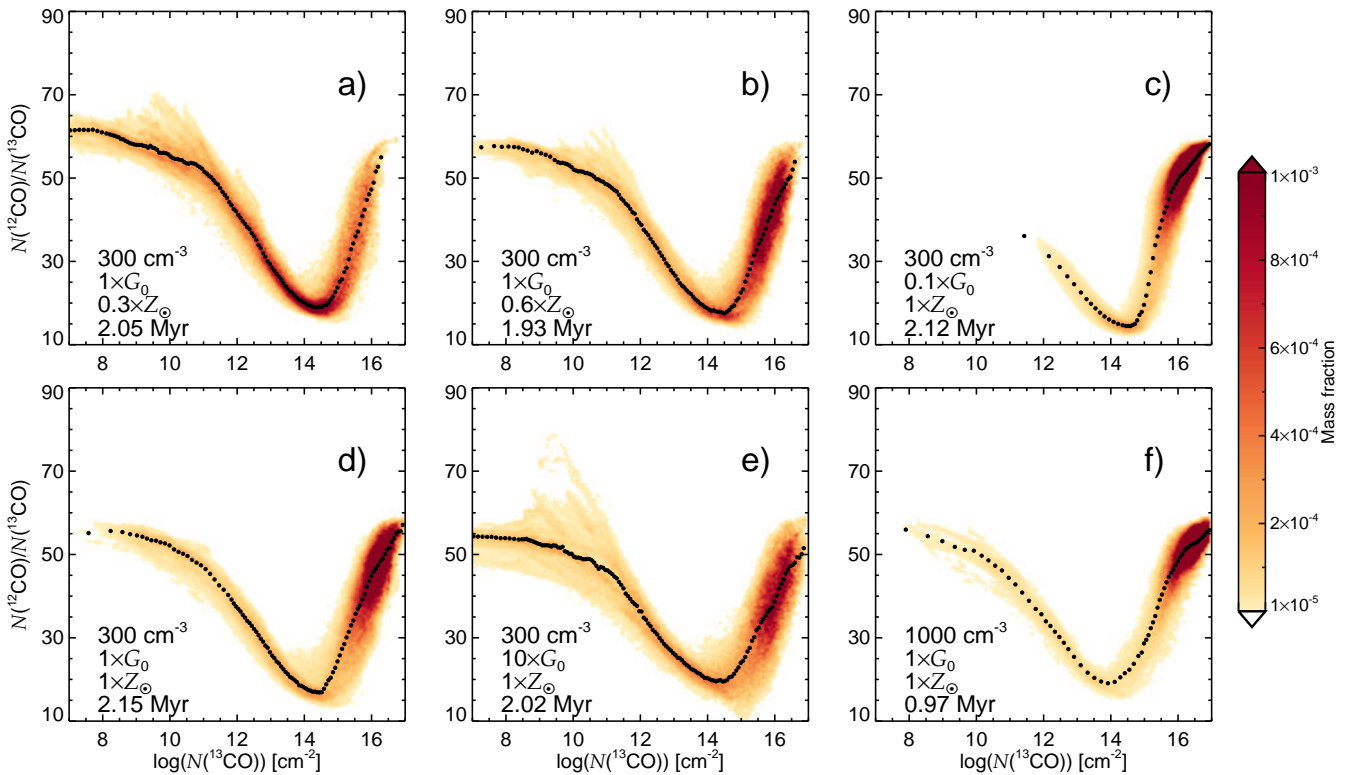}
\caption{CO isotope column density ratio as a function of the $^{13}\textrm{CO}$ column density. Colours and the dashed line have the same meaning as in Fig.~\ref{fig:2d_isorat_12}. The main difference compared to Fig.~\ref{fig:2d_isorat_12} is the steeper slope at column densities larger than the characteristic column density of the dip in the isotope ratio.}
\label{fig:2d_isorat_13}
\end{figure*}

The $^{12}\textrm{CO}/^{13}\textrm{CO}$ isotopic ratio shows a similar correlation with the $^{13}\textrm{CO}$ column density as presented in section~\ref{sec:ratiovsn12co} and a fitting formula could also be derived following the procedure described in section~\ref{sec:fitting} (see Fig.~\ref{fig:2d_isorat_13}). We adopt a similar functional form: $r(N_{13}) = a_{0,13} + a_{1,13} \log_{10}(N_{13}) + a_{2,13} \log_{10}(N_{13})^2 + a_{3,13} \log_{10}(N_{13})^3  + a_{4,13} \log_{10}(N_{13})^4  + a_{5,13} \log_{10}(N_{13})^5$, where $r$ and $N_{13}$ are the isotope ratio and the $^{13}\textrm{CO}$ column density, respectively.  The best fitting coefficients for the individual runs and the combined sample are presented in Table~\ref{table4}. The fitting formula is valid for $^{13}\textrm{CO}$ column densities in the range $3\times10^{10}\,\textrm{cm}^3 - 5\times10^{16}\,\textrm{cm}^3$. We do not consider lines of sight with  column densities less then the lower value and set the isotope ratio to 60 in case of column densities higher than the upper limit. We exclude runs c) and g) from the combined sample.

Table~\ref{12comass} shows the total $^{12}\textrm{CO}$ mass in the simulations, calculated with various methods. In case of the ``real'' mass, we use the self-consistently calculated $^{12}\textrm{CO}$ distribution. In the uniform ratio case we take the $^{13}\textrm{CO}$ column density map and scale it with a uniform $^{12}\textrm{CO}/^{13}\textrm{CO}$ factor of 60. In the case of the fitting formula, we derive $^{13}\textrm{CO}$ column density dependent isotope ratio, and scale $N(^{13}\textrm{CO})$ with the corresponding value. The assumption of a uniform isotope ratio always results in an overestimation of the ``real'' $^{12}\textrm{CO}$ mass. If the fitting formula is applied then the overestimation could be reduced from up to ${\sim}58$ per cent to less than 10 per cent.

This correlation could be used to infer isotope ratios and improve $^{12}\textrm{CO}$ column density estimates based on $^{13}\textrm{CO}$ observations. We note however, that the $^{13}\textrm{CO}$ column density estimation methods are influenced by various sources of errors. At high column densities even the rarer isotopes become optically thick, thus provide only lower limit on the isotope column density. At low column densities, the often assumed local thermodynamic equilibrium is not suitable to calculate molecular level populations precisely, resulting in an underestimation of the column density \citep[e.g.][]{Padoan2000,Molina2014}

\begin{table}
\caption{Total $^{12}\rmn{CO}$ mass derived with various assumptions on the $^{12}\textrm{CO}$ column density (see section~\ref{sec:13coobs}). Positive per cent errors mean overestimation of the self-consistent value. Model g) is discussed in Appendix~\ref{sec:initchemcomp}}
\label{12comass}
\begin{tabular}{ccccccc}
  \hline
  Model & \multicolumn{3}{c}{$^{12}\textrm{CO}$ mass [M$_{\odot}$]} & \multicolumn{2}{c}{Per cent error [\%]} \\
        & ``Real'' & Uniform & Fitting & Uniform & Fitting \\
        &          & ratio    & formula & ratio    & formula \\
  \hline
      a & 0.604  & 0.955 &  0.544  & +58.2 &  -9.9   \\
      b & 3.461  & 5.020 &  3.372  & +45.0 &  -2.6    \\
      c & 18.081  & 21.772 & 16.789   & +20.4 & -7.1     \\
      d & 10.230  & 13.308 & 10.293   & +30.1 & +0.6     \\
      e & 4.275  & 6.144 & 4.684   & +43.7 & -9.6    \\
      f & 18.572  & 21.335 & 19.185   & +14.9 & +3.3     \\
  \hline
      g & 7.050  & 11.534 & 8.745   & +63.6 & +24.0     \\
 \hline
\end{tabular}
\end{table}

\section{The \texorpdfstring{$^{13}\textrm{CO}$}{13CO} line emission} \label{sec:EmissionMaps}

In the previous sections we showed that (photo-)chemical processes have a large impact on the $^{12}$CO/$^{13}\textrm{CO}$ ratio in GMCs and that there is a clear correlation between $N(^{12}\textrm{CO})$, $N(^{13}\textrm{CO})$ and the ratio. We now explore how these effects are reflected in the observable emission line profiles and maps. Could the assumption of uniform isotope ratio still give back approximately the right answer when optical depth effects and detection limits are considered? We try to answer this question quantitatively by comparing line profiles and emission maps produced from self-consistent calculations with those obtained by either adopting a uniform isotope ratio, or one that depends on $N(^{12}{\rm CO})$.

The numerical models are transformed to the observational plane by line radiative transfer modelling. We calculate the emission in a $\pm 6\,\rmn{km\,s}^{-1}$ velocity range around the $J=1\rightarrow0$ transition of  $^{13}\textrm{CO}$ ($\lambda_{0} =$ 2720.41 $\mu$m). For the $^{13}\textrm{CO}$ number density we adopt three distributions: one produced by self-consistent simulations, a second produced by rescaling the $^{12}\textrm{CO}$ distribution with a uniform isotopic ratio of 60 and a third produced by rescaling the $^{12}\textrm{CO}$ distribution using the isotope ratios we obtained from our fitting formula. In the last case, we assume that the isotope ratio is constant along a given line of sight.

In the interstellar medium, the assumption that molecule energy levels are populated according to a thermal distribution (i.e. local thermodynamic equilibrium), is often invalid. To account for non-LTE conditions we use the Large Velocity Gradient (LVG) approximation, described in detail in \citet{Sobolev1957}, \citet{Ossenkopf1997} and \citet{Shetty2011a}. The non-thermal excitation/de-excitation is mainly driven by collisions with other molecules or atoms. As the most abundant particle in the dense ISM, the hydrogen molecule is the most probable collisional partner for CO. \corr{We account for the two spin isomers of the hydrogen molecule and calculate the level population of $^{13}\textrm{CO}$ in two cases: using an ortho-$\textrm{H}_{2}$/para-$\textrm{H}_{2}$ ratio of 3 (the high temperature equilibrium ratio) and of $3 \times 10^{-5}$ (the thermal ratio at 15 K).}
The collisional rates are adopted from the Leiden Atomic and Molecular Database\footnote{\protect\url{http://home.strw.leidenuniv.nl/\~moldata/}} \citep{Schoier2005,Yang2010}.
\corr{In addition to the LVG approximation, in which the photon escape probability in a cell is given by the velocity gradient with the neighbouring cells, we also use the so called ``Doppler catching'' method to correct for numerical artifacts when the Doppler shift between neighbouring cells is larger than the intrinsic line width \citep[][]{Shetty2011b}.}


\begin{table}
\caption{Total $^{13}\rmn{CO}$ mass in the analysed simulation domain when its number density is self-consistently calculated or inferred from $^{12}\rmn{CO}$ with a uniform isotope ratio or using our fitting formula. Negative (positive) per cent errors mean underestimation (overestimation) of the self-consistent value.}
\label{13comass}
\begin{tabular}{ccccccc}
  \hline
  Model & \multicolumn{3}{c}{$^{13}\textrm{CO}$ mass [M$_{\odot}$]} & \multicolumn{2}{c}{Per cent error [\%]} \\
        & ``Real'' & Uniform & Fitting & Uniform & Fitting \\
        &          & ratio    & formula & ratio    & formula \\
  \hline
      a & 0.016  & 0.010 & 0.017   & -36.8 & +6.9    \\
      b & 0.086  & 0.059 & 0.084   & -31.1 & -1.7     \\
      c & 0.374  & 0.310 & 0.386   & -17.0 & +3.3     \\
      d & 0.230  & 0.176 & 0.220   & -23.1 & -3.3     \\
      e & 0.105  & 0.073 & 0.095   & -30.4 & -9.8    \\
      f & 0.366  & 0.319 & 0.358   & -13.0 & -2.2     \\
 \hline
      g & 0.192  & 0.118 & 0.156   & -38.9 & -18.9     \\
 \hline
\end{tabular}
\end{table}

\corr{We use the radiation transfer tool, {\sc radmc-3d}\footnote{\protect\url{http://www.ita.uni-heidelberg.de/\~dullemond/software/radmc-3d/}} \citep{Dullemond} and the physical conditions (number densities, gas temperature, etc.) self-consistently computed in the hydrodynamic simulations to calculate the molecular energy level populations in the above described approximation. Once the level populations are known, the line emission is obtained using the ray-tracing module of {\sc radmc-3d}.} The required input parameters include the number density of the modelled species ($^{13}\textrm{CO}$ in our case), the number density of the collision partners (ortho- and para-hydrogen molecules), the gas temperature, resolved \corr{(i.e. systematic and turbulent)} and unresolved \corr{(``micro-turbulent'')} velocity of the gas, and the line properties (energy levels, statistical weights, Einstein A-coefficients and collision rate coefficients). The SPH data of number densities,  gas velocity and temperature are interpolated to a regular grid of (512 pixel)$^{3}$ as described in Appendix~\ref{sec:grid}. \corr{The ``micro-turbulent'' velocity field, accounting for the small scale velocities unresolved on the grid, is set uniformly according Larson's law by $v_{\textrm{mturb}}=1.1\times(\textrm{0.032 [pc]})^{0.38}=0.297\,\textrm{km s}^{-1}$ \citep{Larson1981}, where 0.032 pc is the linear size of a pixel. Generally, a higher microturbulent velocity results in smoother line profiles. For a detailed discussion on how the choice of microturbulent velocity affects the emission see Appendix A in \citet[][]{Shetty2011b}}. 
The line properties are adopted from \citet{Yang2010}. \corr{The intensity of emission is calculated in position-position-velocity cubes with a velocity resolution of 0.09 km s$^{-1}$}.

We consider three different distributions for the $^{13}\textrm{CO}$ number density when producing the synthetic emission cubes. The first of these is taken directly from our simulation, and hence fully accounts for the effects of chemical fractionation and selective photodissociation. The second distribution is generated by scaling the $^{12}\textrm{CO}$ number densities from the simulation by a uniform factor of $1/60$. Finally, the third distribution is produced by scaling the $^{12}\textrm{CO}$ number densities by a variable factor derived using the isotope ratio-$^{12}\textrm{CO}$ column density fitting function (equation~\ref{eq:fitfun}). In the latter case, we first calculate the $^{12}\textrm{CO}$ column density along the $z$ direction, then use the fitting formula described in section~\ref{sec:fitting} to estimate the average line-of-sight isotope ratio. In the final step, the $^{13}\textrm{CO}$ number density is calculated from the $^{12}\textrm{CO}$ number density and the $^{12}\textrm{CO}$ column density dependent isotope ratio. Table~\ref{13comass} summarizes the total $^{13}\rmn{CO}$ mass for every model and each method used to calculate the $^{13}\rmn{CO}$ number density distribution. These values represent the input number density distributions for the radiation transfer calculations (i.e. no radiation transfer effects are considered). It is clear from the per cent errors that the fitting formula gives a better representation of the ``real'' $^{13}\rmn{CO}$ mass: the uniform scaling of the $^{12}\rmn{CO}$ number density results in an error ranging from \corr{13} to 40 per cent, while when using the fitting formula, the deviation is typically less than \corr{10 per cent}. When adopting a uniform isotope ratio, we always underestimate the total $^{13}\rmn{CO}$ mass. The fitting formula may result in under- or overestimation with comparable absolute errors.

\begin{figure}
\includegraphics[scale=0.68]{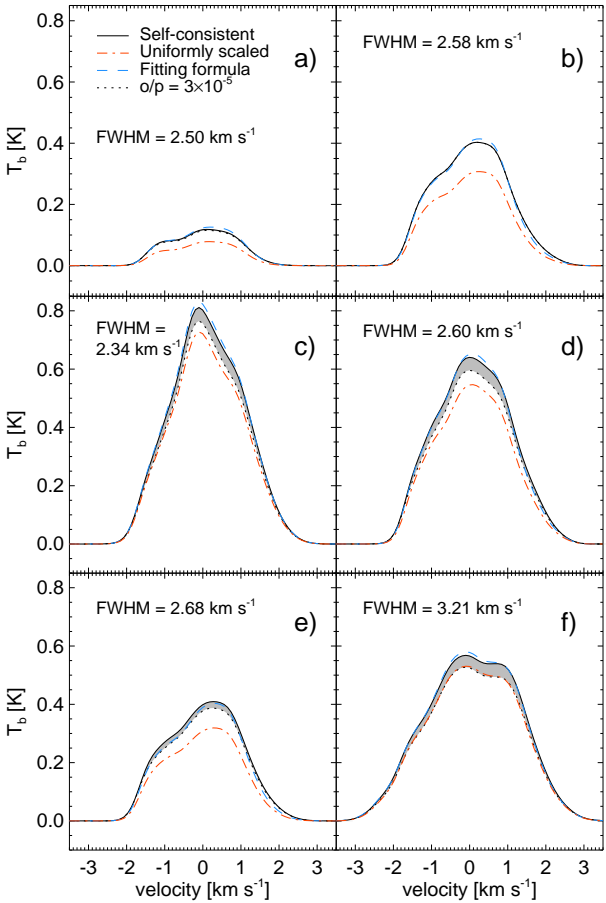}
\caption{\corr{Cloud-averaged emission line profiles of the $^{13}\textrm{CO}$ J=1-0 transition for the six simulations. The solid line shows the self-consistent $^{13}\textrm{CO}$ number density model with the $\textrm{H}_{2}$ ortho-to-para ratio of 3. The dotted line shows the same model, but with $\textrm{ortho-H}_{2}/\textrm{para-H}_{2} = 3\times10^{-5}$ (thermal distribution at 15 K). In reality the ortho-para ratio is between these limiting values, as denoted by the gray shaded area. The dotted-dashed and the dashed lines represents uniformly scaled and fitting formula models, with $\textrm{ortho-H}_{2}/\textrm{para-H}_{2} = 3$. The FWHM of the self-consistent model is indicated on the panels.}}
\label{fig:lines}
\end{figure}

\subsection{Line profiles} \label{sec:lines}
\corr{The results of the radiative transfer calculations are position-position-velocity intensity cubes in units of erg $\rmn{cm}^{-2} \rmn{s}^{-1} \rmn{Hz}^{-1} \rmn{ster}^{-1}$. As the first step the intensity is converted to brightness temperature ($T_{\rmn{b}}$, in K), a quantity usually measured in molecular line observations.}

\corr{The $T_{\rmn{b}}(x,y,v)$ cubes are integrated along the spatial dimensions to obtain the overall $^{13}\rmn{CO}$ J=1-0 line profile of the cloud. The resulting spectra are shown in Fig.~\ref{fig:lines}. For each model, the self-consistent case is calculated using ortho-$\textrm{H}_{2}$/para-$\textrm{H}_{2}$ ratios of 3 and $3 \times 10^{-5}$ for comparison. We find that the choice of the $\rmn{H}_{2}$ spin isomer ratio has less than 10 per cent effect on the integrated intensity of the line and typically a few per cent effect on the line width. Due to the small difference, the uncertain ortho/para ratio in observed clouds \citep{Tielens2013} and that we are focusing on a comparative study, we consider models with the somewhat unrealistic ratio of 3 in the rest of the discussion.}

\corr{We find peak brightness temperatures and full-width-at-half-maximum\footnote{Calculated by the interpolation mode of the IDL procedure  \protect\url{http://www.astro.washington.edu/docs/idl/htmlhelp/library21.html\#fullwid\_halfmax}} (FWHM) values consistent with those observed in molecular clouds with similar masses \citep[e.g. the Perseus molecular cloud complex, see Fig.~4 and Table~2 in][]{Pineda2008}. The peak brightness temperature correlates with the total CO mass in a given simulation (Table~\ref{13comass}). In the case of models a) to e), the line width increases as a larger amount of radiation energy is deposited at greater cloud depths due to less effective shielding (i.e. lower metallicity) or higher ISRF strength. In the case of the high density model (f), the line width is larger due to the initially larger kinetic energy and $\sigma_{\textrm{rms}}$.}

\corr{The line profiles produced by the uniform isotopic ratio and fitting formula models are also shown in Fig.~\ref{fig:lines}. The fitting formula model recovers the self-consistent line shapes and peak brightness temperatures within ${\sim}2$ per cent error, while the choice of a uniform ratio might result in up to 30 per cent error in the peak intensity.}

\corr{The one dimensional velocity dispersion ($\sigma_{1D,m}$) is inferred from the line width assuming a Gaussian line shape by $\sigma_{1D,m} = \textrm{FWHM} / 2.35$. The velocity dispersion calculated this way is always smaller than the real line of sight dispersion, and scarcely depends on the adopted $^{12}\rmn{CO}/^{13}\rmn{CO}$ ratio.}

\begin{table}
\caption{Intrinsic and emission line inferred line of sight velocity dispersions.}
\label{lineprop}
\begin{tabular}{ccccc}
  \hline
  Model & $\sigma_{1D,intrinsic}$ & \multicolumn{3}{c}{$\sigma_{1D,measured}$ [$\textrm{km s}^{-1}$]} \\
        & [$\textrm{km s}^{-1}$] & Self-      & Uniform & Fitting  \\
        & & consistent & ratio    & formula \\

  \hline
      a & 1.39  & 1.06 & 1.03 & 1.04 \\
      b & 1.26  & 1.10 & 1.08 & 1.06 \\
      c & 0.93  & 0.99 & 0.98 & 0.97 \\
      d & 1.09  & 1.11 & 1.08 & 1.07 \\
      e & 1.32  & 1.14 & 1.11 & 1.09 \\
      f & 1.61  & 1.37 & 1.35 & 1.35 \\

 \hline
\end{tabular}
\end{table}

\subsection{Emission maps} \label{sec:compemission}

\begin{figure*}
\includegraphics[scale=0.55]{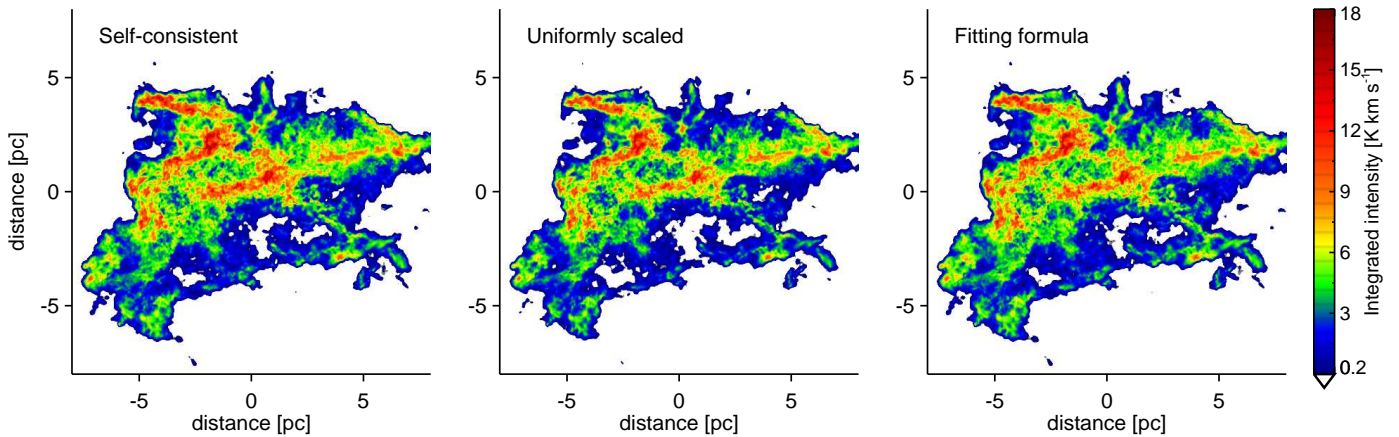}
\caption{Velocity integrated $^{13}\textrm{CO}$ $J=1\rightarrow0$ emission maps derived from run d) in case of the three methods. Pixels in a given map with intensity values less then the chosen detection limit of 0.2 $\textrm{K km s}^{-1}$ are shown in white. The area associated with an emission value is more extended in the self-consistent map than when a uniform $^{12}\textrm{CO}/^{13}\textrm{CO}$ isotopic ratio is assumed. The difference is most prominent at the edge of the cloud (blue region) and in the region of moderate intensity (4-7 $\textrm{K km s}^{-1}$, green and yellow). The latter region contributes the most to the total, spatially and velocity integrated intensity of the cloud. The map produced using our fitting formula agrees well with the self-consistent case.}
\label{fig:intint}
\end{figure*}

The $T_{\rmn{b}}(x,y,v)$ cubes integrated in velocity space, along the line of sight give the zeroth moment maps, $W(^{13}\textrm{CO})$, in units of $\textrm{K\,km\,s}^{-1}$. For the purpose of the comparison we consider pixels with $W(^{13}\textrm{CO})\,>\,0.2\,\textrm{K\,km\,s}^{-1}$ in each map. This limit is comparable to the $0.18\,\textrm{K km s}^{-1}$ $3\sigma$ level of \citet{Goldsmith2008} in the Taurus molecular cloud. This detection limit approximately translates to a $N(^{12}\textrm{CO})$ of $6\times10^{15} \textrm{cm}^{-2}$, suggesting a negligible effect of preferential photodissociation and a more important role of chemical fractionation. We do not consider additional noise in the synthetic maps. See Fig.~\ref{fig:intint} for the zeroth moment maps of run d), calculated with the different assumptions on the $^{12}\rmn{CO}/^{13}\rmn{CO}$ isotope ratio.

\begin{figure*}
\includegraphics[scale=0.468]{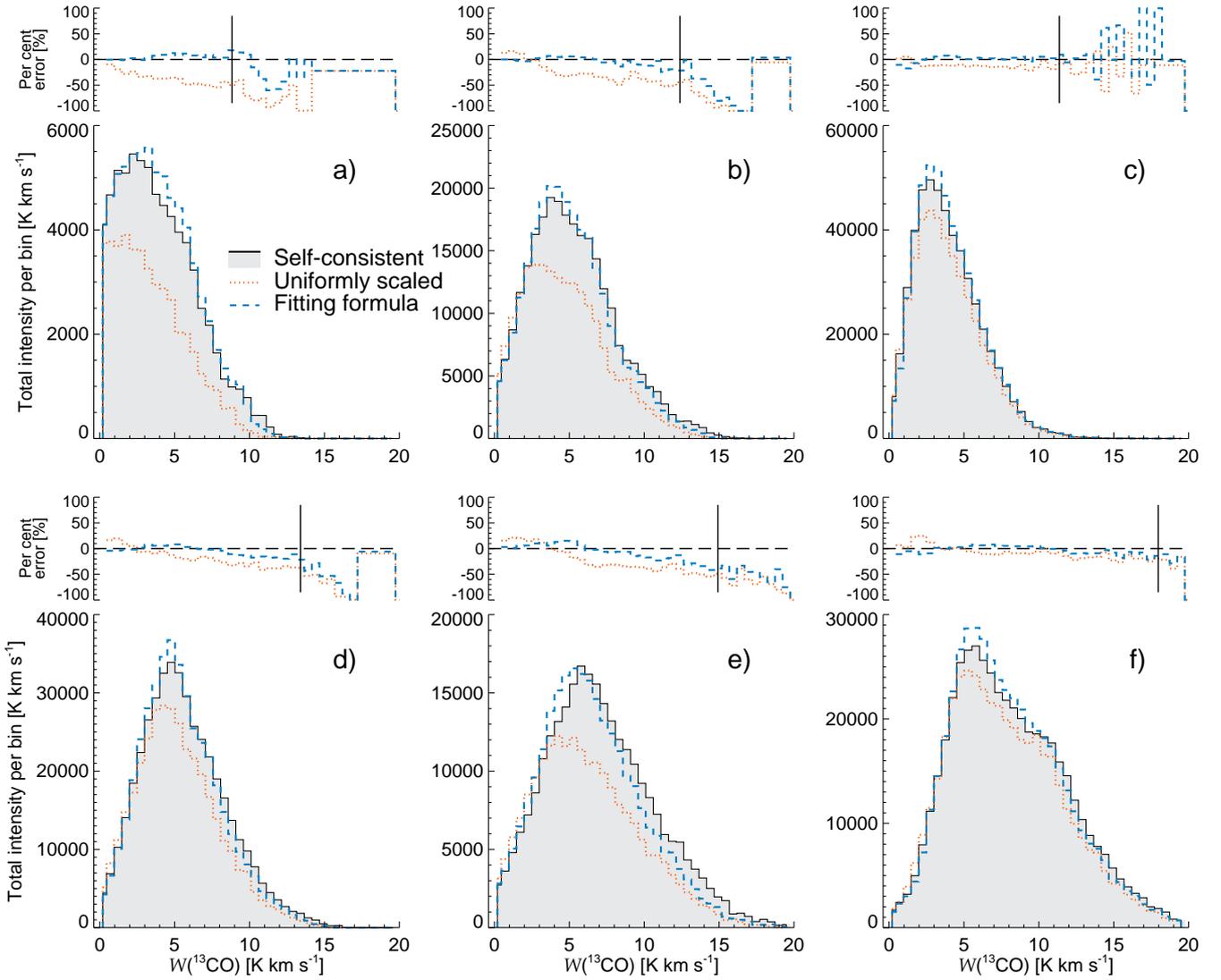}
\caption{Comparison of the $^{13}\rmn{CO}$ ($J=1\rightarrow0$) brightness temperature distributions in the 6 simulations (from a to f). The figure shows which brightness temperatures are contributing the most to the total intensity of the cloud. The filled histograms with the black outlines show the self-consistent models. The orange dotted and blue dashed lines represent the distributions from the uniformly scaled  and fitting formula approximations. The per cent errors of the latter compared to the self-consistent distribution are shown on panels above the intensity distributions. The vertical black line marks the intensity below which each bin contains more than 100 pixels. We only considered pixels with $W(^{13}\textrm{CO}) > 0.2 \textrm{K km s}^{-1}$.}
\label{fig:intensity_dist}
\end{figure*}

Fig.~\ref{fig:intensity_dist} compares the total intensity of pixels falling in a brightness temperature bin (with the bin size of $0.5\,\textrm{K km s}^{-1}$) as a function of brightness temperature for the self-consistent (filled histogram with black outline) and approximate maps (orange dotted and blue dashed for the uniform ratio and fitting formula respectively). The bins with the high total intensity values contribute the most to the total emission. The contribution is determined by the enclosed area of contours associated with the lower and upper boundaries of a bin and the ``mean'' bin intensity. In case of simulation d), for example, the total intensity at high brightness temperatures is low because of the small area occupied by high intensity pixels. At low brightness temperatures, the corresponding area is large, but the mean intensity of pixels is low. The largest intensity contribution comes from pixels with values around $5~\textrm{K km s}^{-1}$, due to the large area occupied and the relatively high mean intensity. The $5~\textrm{K km s}^{-1}$ brightness temperature corresponds to $N(^{13}\textrm{CO}) \approx 1\times10^{16} \textrm{cm}^{-2}$, $N(^{12}\textrm{CO}) \approx 5\times10^{17} \textrm{cm}^{-2}$, and approximately $8.6\times10^{21} \textrm{cm}^{-2}$ total column density (in simulation d). Taking the relatively large area into account, a significant fraction of the $^{13}\textrm{CO}$ and total cloud mass is associated with this intensity range. 

To quantify the deviations form the self-consistent distribution, we plot the per cent errors of the approximate maps on the upper panels of Fig.~\ref{fig:intensity_dist}. At the highest intensities the per cent error shows large fluctuation in both approximate cases. On one hand, the number of pixels in these bins are usually below- or at a few times 10. On the other hand, the $^{12}\textrm{CO}/^{13}\textrm{CO}$ ratio slightly falls below 60 even at the highest intensities/column densities (see Fig.~\ref{fig:fittingformula}). Hereafter we exclude bins containing fewer than 100 pixels (a solid, vertical black lines on the per cent error panels) from the comparison.

Generally, the assumption of uniform isotope ratio results in an up to ${\sim}50$ per cent underestimation at intermediate and high intensities ($W > 5\,\textrm{K km s}^{-1}$) and slight overestimation at low intensities (${\sim}1\,\textrm{K km s}^{-1}$). The fitting formula offers a better fit with an error usually lower than ${\sim}10$ per cent. These errors are comparable to the errors on the scaled (i.e. uniformly or $^{12}\textrm{CO}$ column density dependently) input $^{13}\textrm{CO}$ mass (Table~\ref{13comass}) used in the radiation transport modelling.

The reason for the relatively large difference between the self-consistent and uniformly-scaled maps is that the area associated with a certain $^{13}\textrm{CO}$ column density is systematically underestimated when the uniform isotope ratio is adopted. When the chemical fractionation reaction is taken into account, we find higher $^{13}\textrm{CO}$ number densities at given total and $^{12}\textrm{CO}$ column densities on the intermediate range. In terms of excitation conditions, this means a slight reduction in the number density of collisional partners for the $^{13}\textrm{CO}$ molecules. On the other hand, the kinetic temperature is also slightly higher in these regions. On average, this results in a larger volume with similar excitation conditions and emerging brightness temperature. When self-consistent and uniformly scaled maps are visually compared we find a more extended emission in the former case (see Fig.~\ref{fig:intint}). When using the fitting formula, the extended emission is almost completely recovered.

The error distributions on the upper panels of Fig.~\ref{fig:intensity_dist} show environment dependent trends: in case of simulation c) and f) the per cent error is moderate in both cases (below 10 per cent). In these simulations the isotope-selective reactions play a relatively minor role due to the weak ISRF or the high density (the gas is well shielded close to the cloud edge) and the CO isotope ratio is close to the initial  $^{12}\textrm{C}/^{13}\textrm{C}$ ratio in regions where most of the CO mass dwells (see Fig.~\ref{fig:isoratio}). There is however a qualitative difference between runs c) and f): in the case of run c) the chemical fractionation-dominated region is extended, but due to its low column density, falls below the detection limit. In run f), on the other hand, the corresponding region is very compact.
In simulation a), b), d) and e) the chemical fractionation is important and the fitting formula recovers the self-consistent distribution  better.

We conclude that above the $0.2\,\textrm{K km s}^{-1}$ detection limit only chemical fractionation has a non-negligible effect on the isotope ratio, and therefore on the synthetic $^{13}\textrm{CO}$ emission maps. \corr{The assumption of a uniform isotope ratio in most cases leads to an error as high as 30-40 per cent in the peak brightness temperature and up to 50 per cent in $W(^{13}\textrm{CO})$ at brightness temperatures above ${\sim}5$ $\textrm{K km s}^{-1}$. This approach is a good approximation only when the ISRF is weak, or when the initial density is high. The FWHMs and derived velocity dispersions are insensitive to the adopted isotope ratio. The fitting formula provides a better fit to the self-consistent map in all cases and we recommend its usage in all quantitative investigations.}

\section{Comparison to previous works} \label{sec:com}

\corr{The previous works of \citet{Liszt2007} and \citet{Roellig2013} focus on CO fractionation in photon dominated regions. The former study concludes that the CO isotope ratio in low density PDRs ($n < 10^{2} \textrm{cm}^{-3}$) is determined by the competing processes of selective photodissociation and chemical fractionation and that the ratio does not reflect local properties, such as the CO excitation temperature and gas density. \citet{Roellig2013} find a different behaviour in dense PDRs  ($n > 10^{3} \textrm{cm}^{-3}$) embedded in strong interstellar radiation fields: with the exception of very specific conditions, the $^{12}\textrm{CO}/^{13}\textrm{CO}$ ratio is always smaller than the $^{12}\textrm{C}/^{13}\textrm{C}$ elemental ratio, indicating that isotope-selective photodissociation is much less effective than chemical fractionation.}
The common feature of these studies is that they use simplified geometry and density distributions and scale the radiation field strength and the density to probe a large parameter range. 

\corr{We present cloud simulations with self-consistently calculated density, velocity and temperature structure. Our models correspond to isolated, small ($\textrm{M} \approx 10^{4} \textrm{M}_{\odot}$) molecular clouds before the start of star formation. Thus, they represent an intermediate evolutionary step between the diffuse gas and dense dark cores irradiated by bright young clusters.}

\begin{figure}
\includegraphics[scale=0.325]{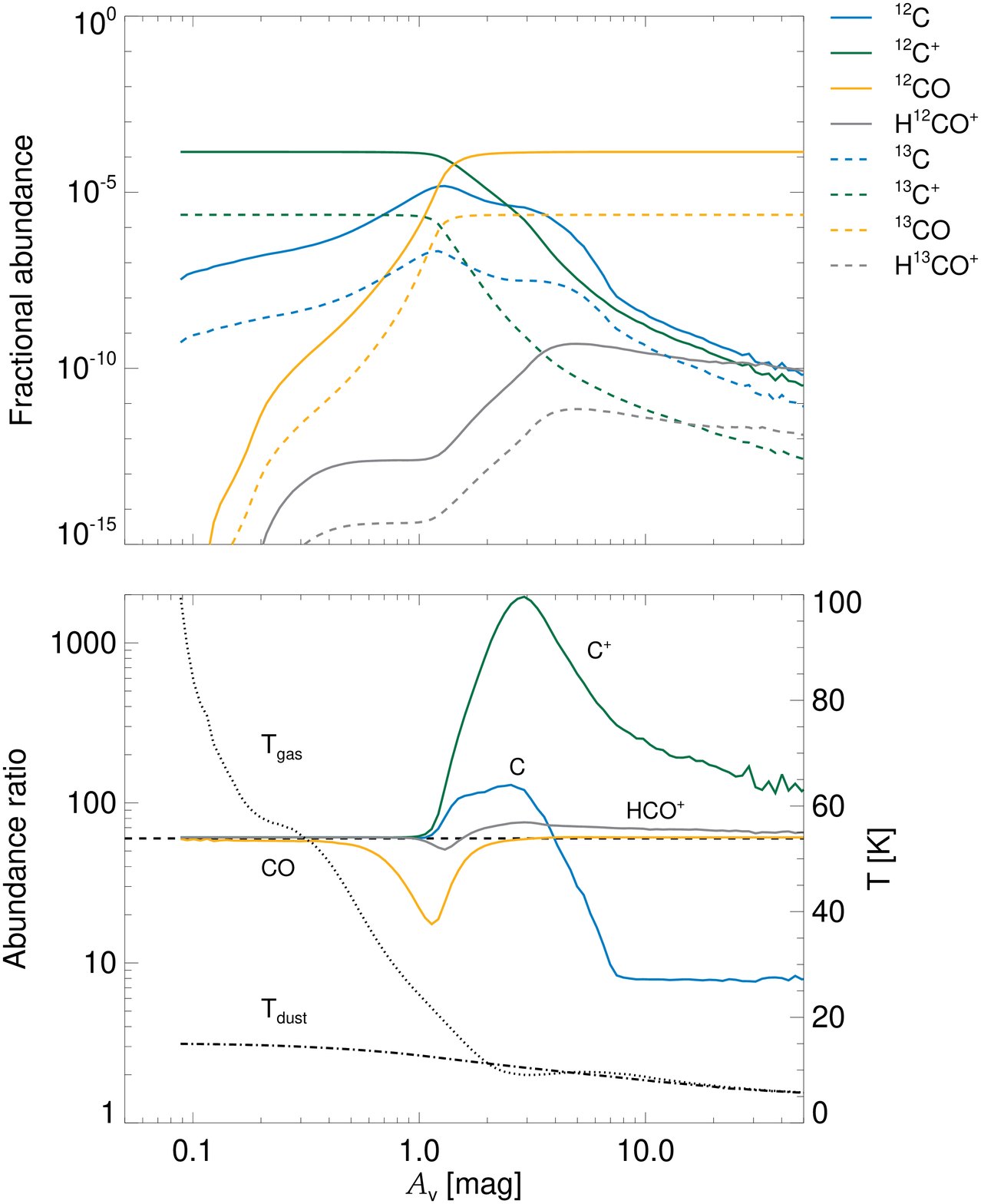}
\caption{The upper panel shows the fractional abundances (relative to total hydrogen nuclei) of the main carbon bearing species and their isotopes as functions of the mean visual extinction in the case of run d). 
The lower panel shows the median isotopic abundance ratios as a function of visual extinction. The dashed line shows the elemental carbon isotope ratio. The dotted and dotted-dashed lines are the median gas and dust temperature at a given visual extinction.}
\label{fig:carb_abuns}
\end{figure}

\corr{The CO isotope ratio in our simulations shows a general behaviour consistent with the results of \citet{Roellig2013}. The upper and lower panels of Fig.~\ref{fig:carb_abuns} present the fractional abundances of carbon bearing species and their isotopic abundance ratios as the function of visual extinction respectively. The figure could be directly compared to Fig.~3 and 4 in \citet{Roellig2013} (note that here we show a cloud with an ISRF 10 times lower than in their figures). At low and high visual extinctions most of the carbon atoms are locked in $\textrm{C}^{+}$ and CO, respectively. The transition of the main isotope is at ${\sim}1.5\,\textrm{mag}$ visual extinction. The conversion of $^{13}\textrm{C}^{+}$ to $^{13}\textrm{CO}$ takes place at a slightly lower visual extinction, due to the Eq.~\ref{eq:fracreact} fractionation reaction becoming the main $^{13}\textrm{CO}$ production pathway \citep{Roellig2013}. Increasing radiation field strength or decreasing metallicity shifts these transitions towards higher $\textrm{A}_{V}$ values. Compared with the \citet{Roellig2013} models, the most important difference is that they find typically an order of magnitude lower $\textrm{C}^{+}$-- CO transitional visual extinction values with similar ISRF strength models. This difference is a result of much higher number densities at the cloud surface, adopted in their models. Their models also produce lower temperatures at similar visual extinction, enhancing the chemical fractionation reaction in those regions. As a result, they find the ``dip'' in the $^{12}\textrm{CO}/^{13}\textrm{CO}$ ratio curve shifted towards lower total/$^{12}\textrm{CO}$ column densities with high end rise coinciding with the UV absorption measurements (see below).}

\corr{We also find regions where the isotope selective photodissociation dominates and the $^{12}\textrm{CO}/^{13}\textrm{CO}$ ratio is increased above the elemental abundance ratio (see yellow zones in Fig.~\ref{fig:isoratio}). Similarly to \citet{Liszt2007}, the typical density is low (around ${\sim}100\, \textrm{cm}^{-3}$) and the temperature is above $40\,\textrm{K}$ (see Fig.~\ref{fig:pdf}). The enhanced CO isotope ratio gas, however is confined to a narrow layer around the denser cloud body, and appears only in the column density ratio maps, when in the given line of sight there is no substantial amount of dense gas.}

On the observational side, \citet{Burgh2007}, \citet{Sheffer2007} and \citet{Sonnentrucker2007} provide ultraviolet absorption measurements of the $\textrm{H}_{2}$, $^{12}\textrm{CO}$, $^{13}\textrm{CO}$ column densities and consequently the $^{12}\textrm{CO} / ^{13}\textrm{CO}$ isotopic ratio. \citet{LisztLucas1998} also determined the CO isotope ratio in nine selected line of sights by measuring millimetre-wavelength absorption and emission. Both the UV absorption and millimetre-wavelength measurements are limited to $^{12}\textrm{CO}$ column densities in the range of $10^{14}\,\textrm{cm}^{-2} < N(^{12}\textrm{CO}) < 4 \times 10^{16}\,\textrm{cm}^{-2}$. At higher column densities the $^{12}\textrm{CO}$ absorption/emission saturates, preventing the direct measurement of its column density. In this high CO column density regime, \citet{Goldsmith2008} fitted the brightness temperature distribution of $^{12}\textrm{CO}$ and $^{13}\textrm{CO}$ emission from the Taurus molecular cloud, using a grid of PDR models with the assumption of an increasing $^{12}\textrm{CO} / ^{13}\textrm{CO}$ abundance ratio with increasing $N(^{12}\textrm{CO})$. Although this method carries significant uncertainties, their results provide valuable constraints on the isotope ratio at high CO column densities.

\begin{figure}
\includegraphics[scale=0.57]{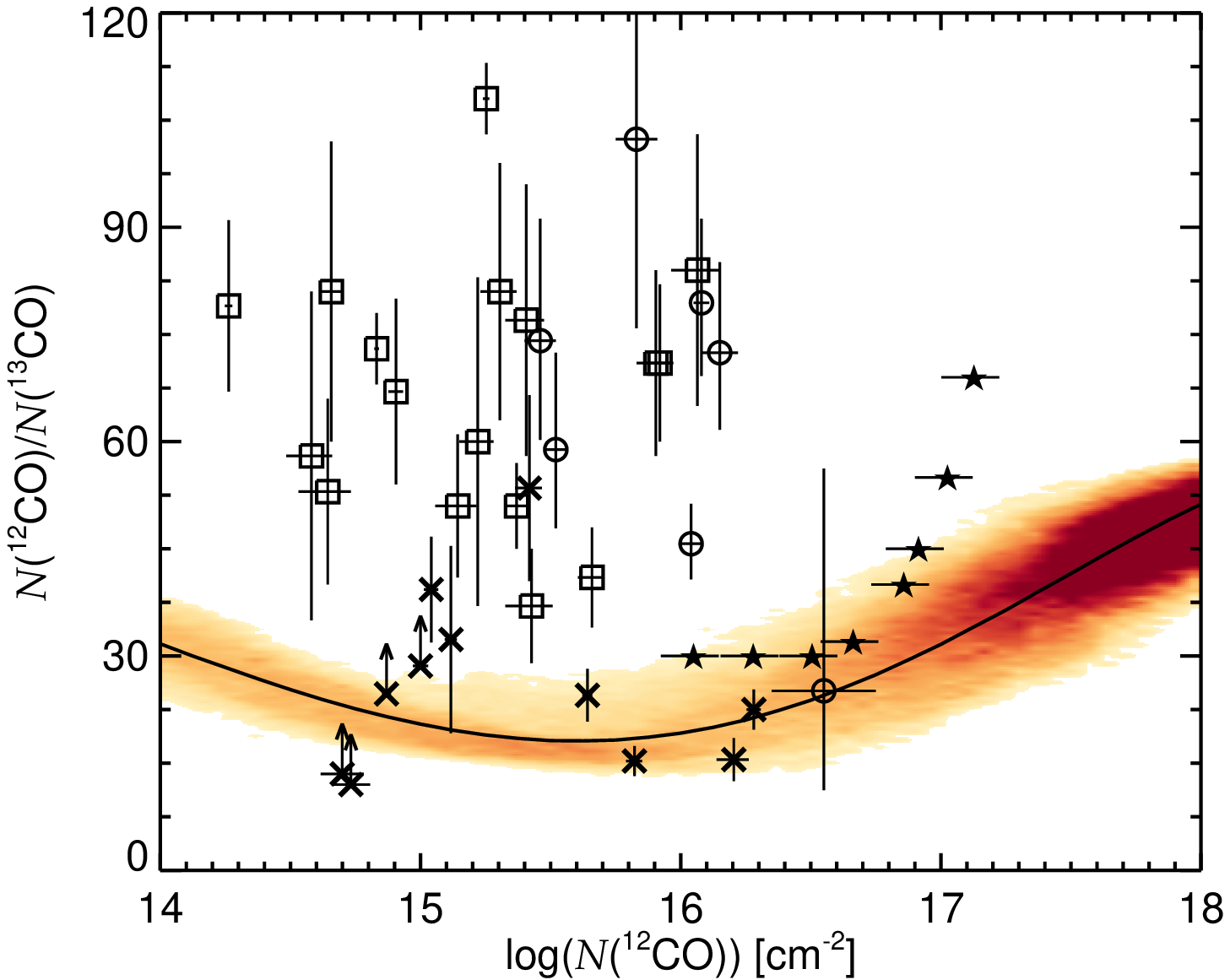}
\caption{$^{12}\textrm{CO}/^{13}\textrm{CO}$ column density ratios from the literature compared to run d) and the adopted fitting formula (black solid line). The ultraviolet absorption measurements are from \citet[][open square]{Sheffer2007} and \citet[][open circle]{Sonnentrucker2007}. The millimetre-wavelength data are from \citet[][``x'' symbol]{LisztLucas1998}. The filled stars are \emph{not direct measurements} but values derived by \corr{emission line} profile fitting presented in \citet{Goldsmith2008}. The colour scale and its meaning is the same as in Fig.~\ref{fig:2d_isorat_12}.
In contrast to the UV absorption measurements, the millimetre-wavelength data and the fitted ratios are largely consistent with the fitting formula and the distribution in the simulation.}
\label{fig:ratio_obs}
\end{figure}

The aforementioned observations are summarized and compared with the fitting formula and run d) in Fig.~\ref{fig:ratio_obs}. The UV measurements are inconsistent with the proposed fitting formula, while the millimetre-wavelength measurements and the fitted ratios are largely consistent with it. \corr{When interpreting UV absorption data one should keep in mind that they preferentially probe low average column density clouds, more consistent with the dilute, moderate temperature PDRs of \citet{Liszt2007}. At the same time, the dense, strongly irradiated PDR models of \citet{Roellig2013} are also partly consistent with the UV measured isotope ratio, due to the relatively cold gas at low column densities, which allows significant chemical fractionation.
On the other hand, these PDR models have more difficulty explaining the low ratios at high $^{12}\textrm{CO}$ column densities, found in millimetre-wavelength measurements. These measurements most likely trace higher average column density clouds similar to the models presented here.}

\section{Summary}  \label{sec:Sum}

We investigate the effects of selective photodissociation and chemical fractionation on the $^{12}\textrm{CO}/^{13}\textrm{CO}$ ratio in realistic hydrodynamical simulations of isolated molecular clouds, considering a range of cloud properties. We aim to quantitatively test the validity of the frequently assumed uniform isotope ratio, when $^{13}\textrm{CO}$ intensities are calculated from hydrodynamical simulations neglecting isotopic chemistry or when $^{12}\textrm{CO}$ column density/mass is inferred from observations of $^{13}\textrm{CO}$ emission.

We find a close correlation between the $^{12}\textrm{CO}$ column density and the isotope ratio, which shows only a weak dependence on cloud conditions within the considered parameter range.
\corr{The isotope-selective-photodissociation is effective and increases the isotope ratio up to 70 in some regions. However, the mass fraction of the gas with higher than elemental ratio is negligible, and has virtually no effect on the column density ratio and on the observables.} The chemical fractionation reaction -- by enhancing the $^{13}\textrm{CO}$ abundance -- reduces the isotope ratio to values as small as 20 in the $10^{15}\,\textrm{cm}^{-2} < N(^{12}\textrm{CO}) < 10^{17}\,\textrm{cm}^{-2}$ range and has a significant effect on the millimetre-wavelength emission. At high CO column densities neither of the isotope-selective reactions are effective, therefore the ratio increases to the value of the elemental isotopic abundance ratio. The isotope ratio varies similarly with the $^{13}\textrm{CO}$ column density.

The correlations depend only weakly on environmental conditions, such as ISRF strength or metallicity. This is because the ability of gas to form CO correlates with its ability to shield itself form the interstellar radiation field. If the irradiation is stronger or the metallicity is lower, the regions of significant $^{12}\textrm{CO}$ and $^{13}\textrm{CO}$ formation shift towards higher total column densities by a similar amount. Furthermore, the characteristic gas temperature of regions where CO reside are very similar in every simulation, providing a consistent condition for the chemical fractionation.

If $^{12}\textrm{CO}$ number densities are provided \citep[e.g. from a simulation of a molecular cloud, such as in][]{GloverClark2012a}, and if we want to calculate the corresponding $^{13}\textrm{CO}$ number densities, then using a uniform $^{12}\textrm{CO}/^{13}\textrm{CO}$ ratio results in the $^{13}\textrm{CO}$ number densities being underestimated by as much as 30-40 per cent (see Table~\ref{13comass}). This can lead to errors of up to 50 per cent in some regions of the derived $^{13}\textrm{CO}$ integrated intensity maps (see Fig.~\ref{fig:intensity_dist}). \corr{The peak brightness temperature is affected by about 30 per cent in this case. However, the adopted isotope ratio has only a few per cent influence on the line width and therefore the derived velocity dispersion. }
In section~\ref{sec:fitting} we derive a fitting formula based on the simulation results to address this issue in a computational cost efficient way (i.e. to infer the correct ratio without the need of the full fractionation chemistry model). When the fitting formula is applied, then the $^{13}\textrm{CO}$ column/number density and emission can be recovered with errors smaller than 10 per cent.

If we have instead $^{13}\textrm{CO}$ column densities (e.g. derived from observations), then using a uniform $^{12}\textrm{CO}/^{13}\textrm{CO}$ ratio to convert them to $^{12}\textrm{CO}$ column densities may result in the overestimation of the $^{12}\textrm{CO}$ column densities by up to 50-60 per cent (see Table~\ref{12comass}). In section~\ref{sec:13coobs} we construct a fitting formula describing the correlation of the isotope ratio and the $^{13}\textrm{CO}$ column density. By applying this formula the errors could be reduced to ${\sim}10$ per cent. 
We note however, that the error introduced by using a fixed $^{12}\textrm{CO}/^{13}\textrm{CO}$ ratio may be smaller then other sources of errors in the determination of the $^{13}\textrm{CO}$ column density (such as the assumption that the excitation temperatures of $^{12}\textrm{CO}$ and $^{13}\textrm{CO}$ are the same and that the $^{13}\textrm{CO}$ emission is optically thin), which can lead to uncertainties of as much as a factor of 4 \citep[see][]{Padoan2000}. Nevertheless, in the $10^{14}\,\textrm{cm}^{-2} < N(^{13}\textrm{CO}) < 10^{16}\,\textrm{cm}^{-2}$ range, where the aforementioned assumptions are valid and the chemical fractionation is effective, the fitting formula provides more accurate conversion to $^{12}\textrm{CO}$ than the assumption of uniform scaling.

The proposed fitting formulae are consistent with millimetre-wavelength $^{12}\textrm{CO} / ^{13}\textrm{CO}$ column density ratio measurements, and underestimate the ratio measured in ultraviolet absorption by a factor of 2-3 (see Fig.~\ref{fig:ratio_obs}). The reason for the discrepancy with the ultraviolet data is probably that the UV measurements are tracing a qualitatively different population of clouds, more consistent with PDR models (diffuse clouds or highly irradiated massive clumps) than with giant molecular clouds.

\corr{Finally, we conclude that the fitting formulae proposed above are good representations of the $^{12}\textrm{CO}/^{13}\textrm{CO}$ isotopic ratio distributions of our hydro-chemical simulations and that they can be used to infer $^{13}\textrm{CO}$ properties from (magneto-) hydrodynamical simulations in a computationally cost-efficient manner, and more precise $^{12}\textrm{CO}$ column density estimates from millimetre-wavelength $^{13}\textrm{CO}$ observations, provided that the molecular clouds under investigation are similar to those presented here.}

\section*{Acknowledgements}
L. Sz. would like to thank M. Sasaki, P. Clark, R. Shetty, D. Semenov, J. Ramsey and V. Gaibler for the useful discussions regarding the work presented in this paper.
The authors would also like to thank the anonymous referee for the suggestions and comments that helped to improve the paper.
The authors acknowledge support from the Deutsche Forschungsgemeinschaft via SFB project 881 ``The Milky Way System'' (sub-projects B2, B3 and B8).
The numerical simulations were partly performed on the KOLOB cluster at the University of Heidelberg and partly on the Milky Way supercomputer, funded by the DFG through Collaborative Research Center (SFB 881) ``The Milky Way System'' (sub-project Z2), hosted and co-funded by the
J\"ulich Supercomputing Center (JSC).
R.~S.~K. acknowledges support from the European Research Council under the European Community's Seventh Framework Programme (FP7/2007-2013) via the ERC Advanced Grant STARLIGHT (project number 339177).

\appendix

\section{Interpolation to a regular grid} \label{sec:grid}
To interpret the results from our Lagrangian hydrodynamic simulations we interpolate the SPH particles to a regular grid. We construct a $512^{3}$ grid centred on the geometrical midpoint of the uniform density sphere defined in the initial conditions. The sides of the grid are 16.2~pc long. The 2D-projected resolution of the grid is therefore $0.032\times0.032~\rmn{pc}^{2}$. This resolution is comparable to the current observationally achievable resolution for nearby star-forming regions in millimetre-wavelength CO line emission \citep[e.g. 0.03~pc linear resolution for Taurus in ][]{Pineda2010}. The interpolated quantities are the total volume density, the gas and dust temperatures, the velocity field and the $\textrm{H}_2$, $^{12}\textrm{CO}$ and $^{13}\textrm{CO}$ number densities.

The density of a voxel (3D pixel) is accumulated from the contribution of SPH particles lying strictly within the voxel and from the contribution of those which have common section with it (i.e. particles with centre coordinates outside of the voxel could contribute). If $r$ is the distance between the centres of the voxel and an SPH particle, the contribution is given by the following smoothing kernel:
\begin{equation}
W(r,h) = \frac{8}{\pi~h^{3}} \times \begin{cases} 1 - 6({\frac{r}{h}})^2 + 6({\frac{r}{h}})^3 
&\mbox{if } 0 \leq \frac{r}{h} \leq 0.5 \\
    2\times(1-{\frac{r}{h}})^3 & \mbox{if } 0.5 < \frac{r}{h} \leq 1 \\                 
    0              & \mbox{if }  \frac{r}{h} > 1\end{cases} 
\label{eq:kernel}
\end{equation}
where $h$ is the smoothing length of the particle defined according equations (5) and (6) in \citet{Springel2005}. The smoothing kernel has units of $\rmn{cm}^{-3}$, and when multiplied with the particle mass gives the contribution of an SPH particle to the voxel density. In case of the other quantities the contribution is the SPH particle value multiplied by the ratio of the density contribution and the total density of the SPH particle.

The interpolation to a regular grid is necessary for the grid based radiation transfer modelling described in section~\ref{sec:EmissionMaps}. For consistency we also analyse the grid interpolated quantities in  section~\ref{sec:ColDenRatio}.

\section{The effect of the initial chemical composition}  \label{sec:initchemcomp}
\corr{In most of the simulations presented in this paper, we adopted initial conditions in which we assumed that the hydrogen is fully molecular (i.e. all of it started in the form of H$_{2}$ and none as atomic hydrogen). In reality, however, molecular clouds form from initially diffuse gas which is a mixture of ionized, neutral and molecular hydrogen. By assuming fully molecular initial conditions we overestimate the $\textrm{H}_{2}$ column density and therefore the $\textrm{H}_{2}$ self- and CO shielding. Self-consistent modelling of the initial chemical composition of clouds, as discussed in e.g. \citet{Clark2012b} and \citet{Smith2014} is out of the scope of this paper, but we investigate the validity of our findings in the other extreme case, when fully atomic initial conditions, with all of the hydrogen in the form of H, are adopted. The simulation with atomic initial composition is model g). All other model parameters have the fiducial value (i.e. as in simulation d). Some of the characteristic quantities of model g) are indicated in Tables~\ref{table3} to \ref{13comass} for comparison.}

\begin{figure}
\includegraphics[scale=0.60]{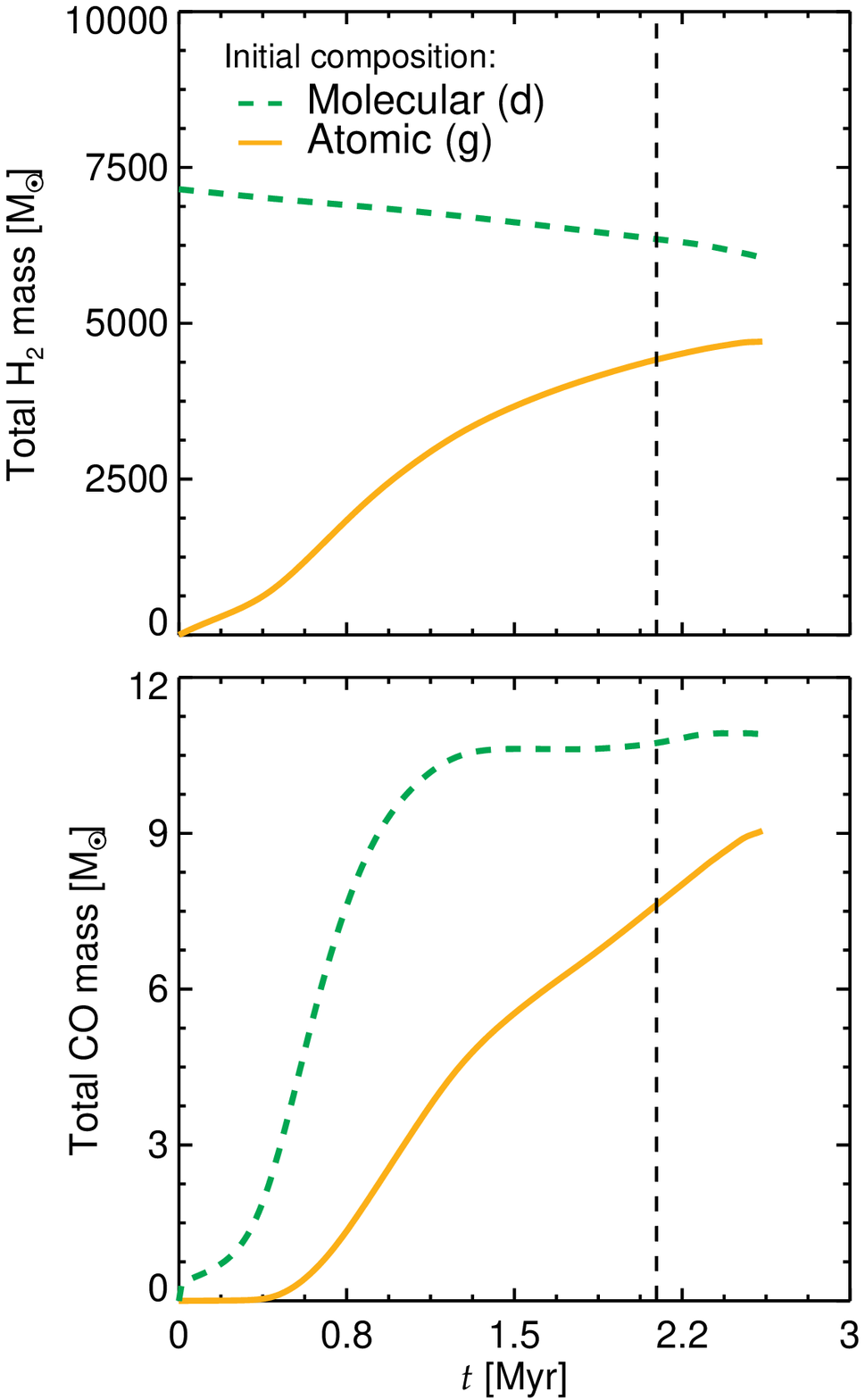}
\caption{\corr{Upper panel: the total $\textrm{H}_{2}$ mass of the cloud as a function of time. The total mass of the cloud is $10^{4} \textrm{M}_{\odot}$. Lower panel: the total CO mass as a function of time. In the fully molecular case the final mass is higher and the formation is faster than in the atomic case. Both simulations are analysed at 2.150  Myr, the time marked by the vertical dashed line.}}
\label{fig:massfrac}
\end{figure}

\corr{The chemical composition of models d) and g) show a very different evolution. The $\textrm{H}_{2}$ mass content (Fig.~\ref{fig:massfrac}) in d) (dashed line) slightly decreases with time due to $\textrm{H}_{2}$ photodissociation and a relatively long $\textrm{H}_{2}$ formation time scale. In the case of g) (solid line) the $\textrm{H}_{2}$ mass gradually increases with time and reaches ${\sim}4800\,\textrm{M}_{\odot}$ (${\sim}70$ per cent of the initial $\textrm{H}_{2}$ mass in d) by the end of the simulation. The CO mass evolution shows similar trends, but in g) it is significantly delayed and the equilibrium CO mass is not reached by the end of the simulation.}

\begin{figure}
\includegraphics[scale=0.60]{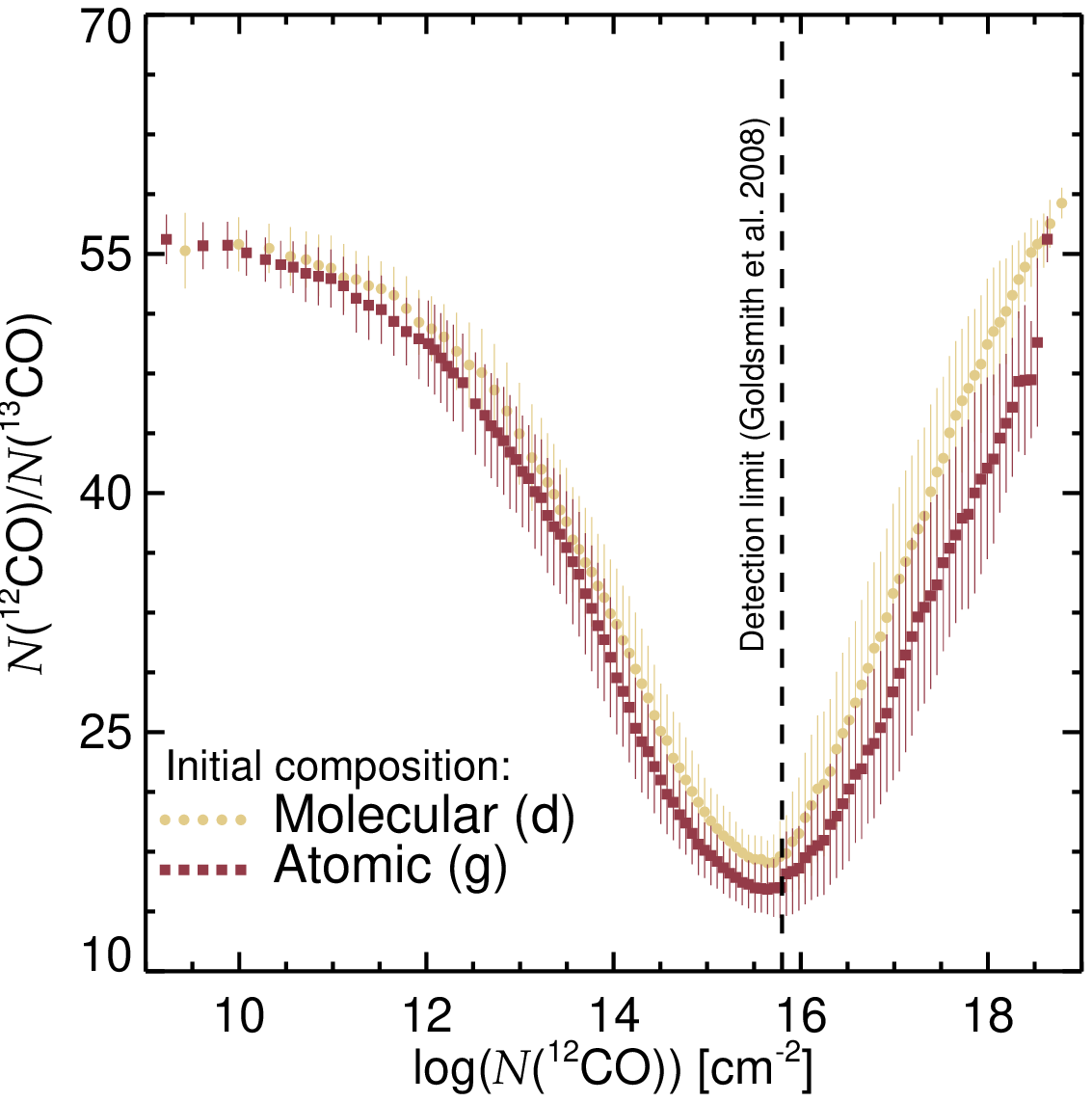}
\caption{\corr{The isotopic ratio as a function of $^{12}\textrm{CO}$ column density in the case of molecular and atomic initial compositions.}}
\label{fig:atomiccurve}
\end{figure}

\corr{Regardless of the lower $\textrm{H}_{2}$ mass and delayed CO formation, the CO isotope ratio - column density relation hold in case of atomic initial conditions (Fig.~\ref{fig:atomiccurve}). The distributions overlap at low column densities, and model g) produces slightly lower isotope ratios at high CO column density. The behaviour of the $^{12}\textrm{CO}/^{13}\textrm{CO}$ ratio in case of d) and g) could be understood as follows. In the simulation with fully atomic initial conditions, there is less shielding of CO by H$_{2}$, owing to the lower molecular content of the cloud, and hence the CO formation threshold shifts to slightly higher densities and extinctions. Lines of sight with column densities $10^{16} < N(^{12}{\rm CO}) < 10^{18} \: {\rm cm^{-2}}$ therefore probe denser regions of the cloud with higher dust extinctions in model g) than in model d). This gas tends to be colder, meaning that the the effect of chemical fractionation is more pronounced, leading to a lower $^{12}\textrm{CO}/^{13}\textrm{CO}$ ratio.}

\begin{thebibliography}{}
\bibitem[\protect\citeauthoryear{Audouze et al.}{1975}]{Audouze1975} 
        Audouze, J., Lequeux, J., \& Vigroux, L., 1975,
        A\&A, 43, 71
\bibitem[\protect\citeauthoryear{Bate et al.}{1995}]{Bate1995} 
        Bate, M. R., Bonnell, I. A., \& Price, N. M., 1995, 
        MNRAS, 277, 362
\bibitem[\protect\citeauthoryear{Beaumont et al.}{2013}]{Beaumont2013}
        Beaumont, C.~N., Offner, S., Shetty, R., Goodman, A., Glover, S.~C.~O., 
         2013, ApJ, 777, 173
\bibitem[\protect\citeauthoryear{Bacmann et al.}{2002}]{Bacmann2002}
	    Bacmann, A., Lefloch, B., Ceccarelli, C., Castets, A., Steinacker, J., Loinard, L.,
	    2002, A\&A, 389L, 6B
\bibitem[\protect\citeauthoryear{Black}{1994}]{Black1994}
        Black, J.~H., 1994, ASP Conf. Ser. 58, in The First Symposium on the Infrared 
        Cirrus and Diffuse Interstellar Clouds, eds. R. M. Cutri \& W. B. Latter, 
        (San Francisco:ASP), 355
\bibitem[\protect\citeauthoryear{Bohlin et al.}{1978}]{Bohlin1978} 
        Bohlin, R. C., Savage, B. D., Drake, J. F., 1978, ApJ, 224, 132
\bibitem[\protect\citeauthoryear{Clark et al.}{2012a}]{Clark2012a} 
        Clark, P. C., Glover, S. C. O., Klessen, R. S., 2012, 
        MNRAS, 420, 745
\bibitem[\protect\citeauthoryear{Clark et al.}{2012b}]{Clark2012b} 
        Clark, P.~C., Glover, S.~C.~O., Klessen, R.~S., Bonnell, I.~A., 2012, 
        MNRAS, 424, 2599        
\bibitem[\protect\citeauthoryear{Burgh et al.}{2007}]{Burgh2007} 
        Burgh, E. B., France, K., McCandliss, S. R., 2007, ApJ, 658, 446
\bibitem[\protect\citeauthoryear{Dullemond}{2012}]{Dullemond}
        Dullemond, C. P. 2012, Astrophysics Source Code Library, 1202.015
\bibitem[\protect\citeauthoryear{Draine}{1978}]{Draine1978}
        Draine, B.~T., 1978, ApJS, 36, 595
\bibitem[\protect\citeauthoryear{Draine \& Bertoldi}{1996}]{DraineBertoldi1996}
        Draine, B.~T.,\& Bertoldi, F., 1996, ApJ, 468, 269
\bibitem[\protect\citeauthoryear{Draine}{2003}]{Draine2003}
        Draine, B.~T., 2003, ApJ, 598, 1026
\bibitem[\protect\citeauthoryear{van Dishoeck \& Black}{1988}]{DishoeckBlack1988} 
        van Dishoek, E.~F., Black, J. H., 1988, ApJ, 334, 771 
\bibitem[\protect\citeauthoryear{Federrath et al.}{2010}]{Federrath2010}
        Federrath, C., Banerjee, R., Clark, P.~C., Klessen, R.~S., 2010
        ApJ, 713, 269
\bibitem[\protect\citeauthoryear{Feldmann et al.}{2012}]{Feldmann2012} 
        Feldmann, R., Gnedin, N.~Y., Kravtsov, A. V., 2012, 
        ApJ, 747, 124
\bibitem[\protect\citeauthoryear{Geiss}{1988}]{Geiss1988}
        Geiss, J., 1988, Reviews in Modern Astronomy 1, ed. G. Klare
        (Heidelberg: Springer-Verlag), pp. 1-27
\bibitem[\protect\citeauthoryear{Goto et al.}{2003}]{Goto2003}
        Goto, M., Usuda, T., Takato, N., Hayashi, M., Sakamoto, S., Gaessler, W.,
        Hayano, Y., et al., 2003, ApJ, 598, 1038
\bibitem[\protect\citeauthoryear{Glover \& Mac Low}{2007}]{GloverMacLow2007} 
        Glover, S. C.~O., \& Mac Low, M.-M. 2007, ApJ, 659, 1317
\bibitem[\protect\citeauthoryear{Glover et al.}{2010}]{Glover2010} 
        Glover, S. C.~O., Federrath, C., Mac Low, M.-M., \& Klessen, 
        R. S., 2010, MNRAS, 404, 2
\bibitem[\protect\citeauthoryear{Glover \& Mac Low}{2011}]{GloverMacLow2011}
        Glover, S. C. O., {Mac Low}, M.-M., 2011, MNRAS, 412, 337
\bibitem[\protect\citeauthoryear{Glover \& Clark}{2012a}]{GloverClark2012a} 
        Glover, S.~C.~O., Clark, P. C., 2012a, MNRAS, 421, 9
\bibitem[\protect\citeauthoryear{Glover \& Clark}{2012b}]{GloverClark2012b} 
        Glover, S.~C.~O., Clark, P. C., 2012b, MNRAS, 421, 116
\bibitem[\protect\citeauthoryear{Goldsmith et al.}{2008}]{Goldsmith2008} 
        Goldsmith, P. F., Heyer, M., Narayanan, G., Snell, R., 
        Li, D., Brunt, C., 2008, ApJ, 680, 428
\bibitem[\protect\citeauthoryear{G\'{o}rski et al.}{2005}]{Gorski2005}
        G\'{o}rski K. M., Hivon E., Banday A. J., Wandelt B. D., Hansen F. K., Reinecke M., Bartelmann M., 
        2005, ApJ, 622, 759
\bibitem[\protect\citeauthoryear{Gredel et al.}{1987}]{Gredel1987}
        Gredel, L., Lepp, S., Dalgarno, A., 1987, ApJ, 323, L137
\bibitem[\protect\citeauthoryear{Habing}{1968}]{Habing1968}
        Habing, H.~J., 1968, Bull. Astron. Inst. Netherlands, 19, 421
\bibitem[\protect\citeauthoryear{Hockney \& Eastwood}{1988}]{HockneyEastwood1988}
        Hockney, R. W., Eastwood, J. W., Computer Simulation 
        Using Particles (Institute of Physics, Bristol, England, 1988)        
\bibitem[\protect\citeauthoryear{Jappsen et al.}{2005}]{Jappsen2005}
        Jappsen, A.-K., Klessen, R. S., Larson, R. B., Li, Y., Mac Low, M.-M., 
        2005, A\&A, 435, 611
\bibitem[\protect\citeauthoryear{Langer}{1976}]{Langer1976} 
        Langer, W.~D., 1977, ApJ, 212, 39
\bibitem[\protect\citeauthoryear{Langer \& Penzias}{1990}]{LangerPenzias1990} 
        Langer, W.~D., Penzias, A.~A., 1990, ApJ, 357, 477
\bibitem[\protect\citeauthoryear{Larson}{1981}]{Larson1981} 
        Larson, R.~B., 1981, MNRAS, 194, 809
\bibitem[\protect\citeauthoryear{Lee et al.}{1996}]{Lee1996}
        Lee, H.-H., Herbst, E., Pineau des Forets, G., Roueff, E., 
        \& Le Bourlot, J. 1996, A\&A, 311, 690
\bibitem[\protect\citeauthoryear{Leroy et al.}{2011}]{Leroy2011}
        Leroy, A. K., Bolatto, A, Gordon, K., Sandstrom, K., et al., 
        2011, ApJ, 737, 12
\bibitem[\protect\citeauthoryear{Liszt \& Lucas}{1998}]{LisztLucas1998}
        Liszt, H. S., Lucas, R., 1998, A\&A, 339, 561        
\bibitem[\protect\citeauthoryear{Liszt}{2007}]{Liszt2007}
        Liszt, H. S., 2007, A\&A, 476, 291        
\bibitem[\protect\citeauthoryear{Lucas \& Liszt}{1998}]{LucasLiszt1998}
        Lucas, R., Liszt, H. S., 1998, A\&A, 337, 246
\bibitem[\protect\citeauthoryear{Mac Low et al.}{1998}]{MacLow1998}
        Mac Low, M.-M., Klessen, R. S., Burkert, A., Smith, M. D., 1998, 
        PhRvL, 80, 2754
\bibitem[\protect\citeauthoryear{Maloney et al.}{1996}]{Maloney1996}
        Maloney, P. R., Hollenbach, D. J., Tielens, A. G. G. M., 1996, 
        ApJ, 466, 561
\bibitem[\protect\citeauthoryear{Mitchell \& Maillard}{1993}]{MitchellMaillard1993}
        Mitchell, G.~F., \& Maillard, J.-P., 1993, ApJ, 404, L79
\bibitem[\protect\citeauthoryear{Molina et al.}{2014}]{Molina2014} 
        Molina, F., Glover, S. C. O., Shetty, R., Klessen, R. S., 
        2014, MNRAS, in prep.
\bibitem[\protect\citeauthoryear{Nelson \& Langer}{1999}]{NelsonLanger99} 
        Nelson R.~P., Langer, W.~D, 1999, ApJ, 524, 923 
\bibitem[\protect\citeauthoryear{Ossenkopf}{1997}]{Ossenkopf1997} 
        Ossenkopf, V., 1997, New Astronomy, 2, 365
\bibitem[\protect\citeauthoryear{Padoan et al.}{2000}]{Padoan2000}
        Padoan, P., Juvela, M., Bally, J., \& Nordlund,  \AA{}., 2000, ApJ, 529, 259
\bibitem[\protect\citeauthoryear{Pineda et al.}{2008}]{Pineda2008} 
        Pineda, J. E., Caselli, P., Goodman, A. A., 2008, ApJ, 679, 481
\bibitem[\protect\citeauthoryear{Pineda et al.}{2010}]{Pineda2010} 
        Pineda, J. L., Goldsmith, P. F., Chapman, N., Snell, 
        R. L., Li, D., Cambr\'{e}sy, L., Brunt, C., 2010, 
        ApJ, 721, 686
\bibitem[\protect\citeauthoryear{Prasad \& Tarafdar}{1983}]{PrasadTarafdar1983}
        Prasad, S. S., Tarafdar, S. P., 1983, ApJ, 267, 603
\bibitem[\protect\citeauthoryear{R\"{o}llig \& Ossenkopf}{2013}]{Roellig2013}
        R\"{o}llig, M., Ossenkopf, V., 2013,  A\&A, 550, 56 
\bibitem[\protect\citeauthoryear{Sch\"{o}ier et al.}{2005}]{Schoier2005} 
        Sch\"{o}ier, F. L., van der Tak, F. F. S., van Dishoeck, 
        E. F., Black, J. H., 2005, A\&A, 432, 369
\bibitem[\protect\citeauthoryear{Scoville et al.}{1983}]{Scoville1983}
        Scoville, N., Klienmann, S.~G., Hall, D.~N.~B., \& Ridgway, 
        S.~T., 1983, ApJ, 275, 201
\bibitem[\protect\citeauthoryear{Sembach et al.}{2000}]{Sembach2000} 
        Sembach, K. R., Howk, J. C., Ryans, R. S. I., 
        \& Keenan, F. P., 2000, ApJ, 528, 310
\bibitem[\protect\citeauthoryear{Sheffer et al.}{2007}]{Sheffer2007} 
        Sheffer, Y., Rogers, M., Federman, S. R., Lambert, D. L., Gredel, R.,
        2007, ApJ, 667, 1002
\bibitem[\protect\citeauthoryear{Shetty et al.}{2011a}]{Shetty2011a} 
        Shetty, R., Glover, S. C., Dullemond, C. P., Klessen, 
        R. S., 2011a, MNRAS, 412, 1686
\bibitem[\protect\citeauthoryear{Shetty et al.}{2011b}]{Shetty2011b} 
        Shetty, R., Glover, S. C., Dullemond, C. P., Ostriker, 
        E. C., Harris, A. I., Klessen, R. S., 2011b, MNRAS, 415, 3253
\bibitem[\protect\citeauthoryear{Smith \& Adams}{1980}]{SmithAdams1980} 
        Smith, D., Adams, N. G., 1980, ApJ, 242, 424
\bibitem[\protect\citeauthoryear{Smith et al.}{2014}]{Smith2014} 
        Smith, R. J., Glover, S. C. O., Clark, P. C., Klessen, R. S., Springel, V.,
        2014, MNRAS, 441, 1628
\bibitem[\protect\citeauthoryear{Sobolev}{1957}]{Sobolev1957}
        Sobolev, V. V., 1957, SvA, 1, 678
\bibitem[\protect\citeauthoryear{Sonnentrucker et al.}{2007}]{Sonnentrucker2007} 
        Sonnentrucker, P., Welty, D. E., Thorburn, J. A., York, D. G., 
        2007, ApJS, 168, 58
\bibitem[\protect\citeauthoryear{Springel}{2005}]{Springel2005} 
        Springel, V., 2005, MNRAS, 364, 1105
\bibitem[\protect\citeauthoryear{Tafalla}{2004}]{Tafalla2004}
        Tafalla, M., Myers, P. C., Caselli, P., Walmsley, C. M., 2004, A\&A, 416, 191
\bibitem[\protect\citeauthoryear{Tielens}{2013}]{Tielens2013} 
        Tielens, A. G. G. M., 2013, Rev. Mod. Phys., 85, 1021
\bibitem[\protect\citeauthoryear{Visser et al.}{2009}]{Visser2009} 
        Visser, R., van Dishoeck, E. F., Black, J. H., 2009, A\&A, 503, 323
\bibitem[\protect\citeauthoryear{Watson et al.}{1976}]{Watson1976} 
        Watson, W. D., Anicich, V. G., Huntress, W. T., 
        1976, ApJ, 205, 165
\bibitem[\protect\citeauthoryear{Williams et al.}{2011}]{Williams2011}
        Williams, T., Kelley, C., et al., 
        2011, Gnuplot 4.5: an interactive plotting program
\bibitem[\protect\citeauthoryear{Wilson}{1999}]{Wilson1999} 
        Wilson, T. L., 1999, Rep. Prog. Phys., 62, 143
\bibitem[\protect\citeauthoryear{Wilson}{2009}]{Wilson2009} 
        Wilson, T. L., Rohlfs, K., H\"{u}ttemeister, S., 2009, 
        in Tools of Radio Astronomy, Springer
\bibitem[\protect\citeauthoryear{Wolfire}{2010}]{Wolfire2010}        
        Wolfire, M. G., Hollenbach, D., McKee, C. F., 2010, ApJ, 716, 1191
\bibitem[\protect\citeauthoryear{Yang et al.}{2010}]{Yang2010} 
        Yang, B., Stancil, P.C., Balakrishnan, N., Forrey, 
        R. C., 2010, ApJ, 718, 1062

\end{thebibliography}
\end{document}